\documentclass[12pt]{iopart}
\usepackage{iopams}
\usepackage[T1]{fontenc}
\usepackage[latin9]{inputenc}
\usepackage{float}
\expandafter\let\csname equation*\endcsname\relax
\expandafter\let\csname endequation*\endcsname\relax
\usepackage{amsmath}
\usepackage{amssymb}
\usepackage{graphicx}
\usepackage{esint}
\usepackage{subfigure}
\usepackage{setspace}

\begin{document}

\title{Quantum Storage and Retrieval of Light by Sweeping the Atomic Frequency}

\author{H. Kaviani$^\ast$, M. Khazali$^\ast$, R. Ghobadi, E. Zahedinejad, K. Heshami and C. Simon}

\address{Institute for Quantum Science and Technology, Department of Physics and Astronomy, University of Calgary, Alberta, Canada T2N 1N4}
\address{$^\ast$ These authors contributed equally to this work}
\begin{abstract}

We propose a quantum memory protocol based on dynamically changing the resonance frequency of an ensemble of two-level atoms. By sweeping the atomic frequency in an adiabatic fashion, photons are reversibly transferred into atomic coherences. We present a polaritonic description for this type of storage, which shares some similarities with Electromagnetically Induced Transparency (EIT) based quantum memories. On the other hand the proposed memory is also linked to the Gradient Echo Memory (GEM) due to the effective spatial gradient that pulses experience in the medium. We discuss a possible implementation of the protocol in hollow-core photonic crystal fibers.

\end{abstract}

\maketitle

\section{Introduction}
Storage and retrieval of photons on demand is essential for quantum information processing tasks such as long-distance quantum communication and distributed quantum computing \cite{Optical_Quantum_Memory,Review_Sorenson,Christoph_Review}. Quantum memories for photons can be realized by coherent control of the atom-photon interaction. Some protocols for quantum memory introduce a control beam in a three-level atomic configuration to manipulate the interaction between the signal pulse and the atoms (e.g. Electrically Induced Transparency (EIT) \cite{Quantum_memory_Dark_state} and Raman schemes \cite{Raman_Nunn,Gorshkov,Raman_nature}). Other protocols use the phenomenon of photon echo \cite{Photon_Echo_1,Photon_Echo_2} to achieve controlled atom-photon interaction. In this type of memory the atom-photon interaction is controlled in a more indirect way by a dephasing-rephasing process owing to the inhomogeneous broadening of the medium. Atomic Frequency Comb memory (AFC) \cite{AFC}, Controlled Reversible Inhomogeneous Broadening memory (CRIB) \cite{CRIB} and Gradient Echo Memory (GEM) \cite{GEM_Analytical} are examples of this type of protocols. In contrast to EIT and Raman protocols, which require at least three atomic levels, photon echo based protocols can be implemented using just two atomic levels, relying on long excited state lifetimes for storage. Such long lifetimes for optical transitions are common for example in rare-earth ion doped crystals.

Recently it has been shown that some of these protocols can be emulated by dynamically changing certain characteristics of an ensemble of two-level atoms. For instance, it has been shown that by dynamically controlling the transition dipole moment of an ensemble of two-level atoms, one can emulate Raman-type quantum memories \cite{Controllable_Dipole}. More recently it has been shown that changing the refractive index of the host medium of two-level atoms is equivalent to the GEM memory \cite{n_of_t}.

Here we study another way of manipulating the atom-photon interaction, namely by sweeping the atomic frequency of a homogeneously broadened ensemble of two-level atoms in time. The energy levels of the two-level atoms can be changed by applying a magnetic or electric field depending on the system.

We demonstrate that the proposed ``Atomic Frequency Sweep (AFS)'' protocol can be described in terms of polaritons, which share some similarities with the dark-state polariton in EIT \cite{Polariton}. By changing the detuning, the pulse slows down and is stored in atomic coherences. Changing the detuning in AFS plays the role of changing the control field in EIT. However there are also differences between AFS and EIT based protocols, in particular, in our protocol the pulse does not shrink when it enters the medium, in contrast with EIT. Moreover here the polaritons are associated with the optical transition.

On the other hand, the AFS protocol is similar to the GEM protocol for the regime in which pulses are short compared to the medium length. While we change the detuning in time during the propagation of the pulse through the medium, the pulse effectively sees a spatial gradient in the energy levels of the atoms, and the excited atomic coherences become dephased.

Due to its similarities with both EIT and GEM, the AFS protocol thus constitutes a bridge between protocols based on coherent control and those based on photon echo.

This paper is organized as follows. In section $2$ we describe the scheme using a polaritonic description. In section $3$ we discuss the experimental requirements and a possible implementation. In section 4 we give our conclusions.

\section{AFS Quantum Memory: Polaritonic Description}

The Maxwell-Bloch equations for the field and atomic polarization (or, equivalently, the single-excitation photonic and atomic wavefunctions) under the usual dipole, rotating-wave and slowly varying envelope approximations are \cite{Review_Sorenson,Gorshkov,Controllable_Dipole}

\begin{equation}
\frac{\partial}{\partial t}\sigma_{ge}(z,t)=-i(\Delta(t)-i\gamma)\sigma_{ge}+i\beta\mathcal{E}(z,t),
\label{field_equation}
\end{equation}

\begin{equation}
(c\frac{\partial}{\partial z}+\frac{\partial}{\partial t}){\cal E}(z,t)=i\beta\sigma_{ge},\label{polarization_equation}
\end{equation}

where $\Delta(t)$ is the detuning, $\gamma$ is decay rate and $\beta=g\sqrt{N}$ is
the collective coupling constant, with $g$ the single-atom coupling and $N$ the number of atoms.

Our proposed scheme is best described in a polaritonic picture. We can write
Eq. (\ref{field_equation}) and Eq. (\ref{polarization_equation}) in k-space as

\begin{equation}
\frac{\partial}{\partial t}\begin{pmatrix}\mathcal{E}(k,t)\\
\sigma_{ge}(k,t)
\end{pmatrix}=i\begin{pmatrix}-kc & \beta\\
\beta & -\Delta(t)
\end{pmatrix}\begin{pmatrix}\mathcal{E}(k,t)\\
\sigma_{ge}(k,t)
\end{pmatrix}.
\end{equation}

For the sake of simplicity, we neglect the decay rate $\gamma$ for now. We will discuss its effects at the end of this section.

One can solve this set of equations by looking at its eigenmodes:

\begin{equation}
\Psi(k,t)=\cos\theta(k)\mathcal{E}(k,t)+\sin\theta(k)\sigma_{ge}(k,t),
\label{Psi_Polariton}
\end{equation}

\begin{equation}
\Phi(k,t)=-\sin\theta(k)\mathcal{E}(k,t)+\cos\theta(k)\sigma_{ge}(k,t),
\label{Phi_Polariton}
\end{equation}

where the mixing angle $\theta(k)$ is given by

\begin{equation}
\sin2\theta(k)=\frac{2\beta}{\sqrt{4\beta^{2}+(ck+\Delta(t))^{2}}},\label{eq:10}
\end{equation}

\begin{equation}
\cos2\theta(k)=\frac{-(ck+\Delta)}{\sqrt{4\beta^{2}+(ck+\Delta(t))^{2}}}.\label{eq:11}
\end{equation}

In terms of these eigenmodes, the equations of motion become

\begin{equation}
\frac{\partial\Psi(k,t)}{\partial t}-i\lambda_{1}(k)\Psi(k,t)=\dot{\theta}(k)\Phi(k,t),\label{eq:12}
\end{equation}

\begin{equation}
\frac{\partial\Phi(k,t)}{\partial t}-i\lambda_{2}(k)\Phi(k,t)=\dot{\theta}(k)\Psi(k,t),
\label{eq:13}
\end{equation}
where $\lambda_{1}=\beta \cot \theta$ and $\lambda_{2}=-\beta \tan \theta$ are the eigenvalues of the system
and $\dot{\theta}=-\frac{\dot{\Delta}}{4\beta}\sin^{2}2\theta\label{eq:16}$ is the time derivative of the mixing angle.

Assuming the adiabatic condition $\dot{\theta} \ll \lambda_{1},\lambda_{2}$ holds during the storage and retrieval process, we can neglect $\dot{\theta}$ compared to $\lambda_{1}$ and $\lambda_{2}$ in Eq. (\ref{eq:12}) and Eq. (\ref{eq:13}) so that the eigenmodes become decoupled.

Now we Taylor-expand the eigenvalues in terms of $k$. In the regime where the coupling constant is larger than the bandwidth of the input pulse ($\beta \gg \Delta \omega$), we can keep terms up to first order in $k$ and neglect higher orders of $k$. With that, and transforming the equations back to $z$ space, we find the equations of motion for the eigenmodes $\Psi$ and $\Phi$ in real space:

\begin{equation}
\frac{\partial\Psi(z,t)}{\partial t}+c\cos^{2}\theta(0)\frac{\partial\Psi(z,t)}{\partial z}=i\lambda_{1}(0)\Psi(z,t),
\label{Polariton_real_1}
\end{equation}
\begin{equation}
\frac{\partial\Phi(z,t)}{\partial t}+c\sin^{2}\theta(0)\frac{\partial\Phi(z,t)}{\partial z}=i\lambda_{2}(0)\Phi(z,t).
\label{Polariton_real_2}
\end{equation}

\begin{figure}[H]
\centering
\includegraphics[scale=0.7]{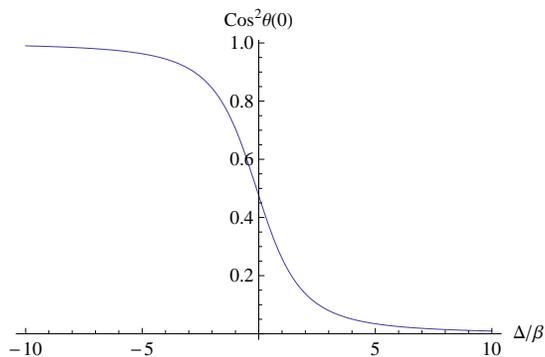}
\caption{Mixing angle $\theta(0)$ as a function of detuning.}
\label{fig:theta}
\end{figure}

Eq. (\ref{Polariton_real_1}) and Eq. (\ref{Polariton_real_2}) indicate that polaritons $\Psi$ and $\Phi$ travel with group velocity $v_{g}=c\cos^{2} \theta(0)$ and $v_{g}=c \sin^{2} \theta(0)$ respectively. Figure \ref{fig:theta} shows how the mixing angle $\theta(0)$ varies with the detuning $\Delta$.

Therefore if we start from a large negative (positive) detuning $-\Delta_{0}$ ($+\Delta_{0}$) compared to the coupling, we can couple the input pulse to the polariton $\Psi$ ($\Phi$). By sweeping the detuning adiabatically to a large positive (negative) $+\Delta_{0}$ ($-\Delta_{0}$), we can slow down the light and convert it to the atomic coherence reversibly. The detuning thus plays a role that is analogous to that of the Rabi frequency of the coupling field in the case of EIT.

\begin{figure}[H]
\centering
\subfigure{\includegraphics[scale=0.4]{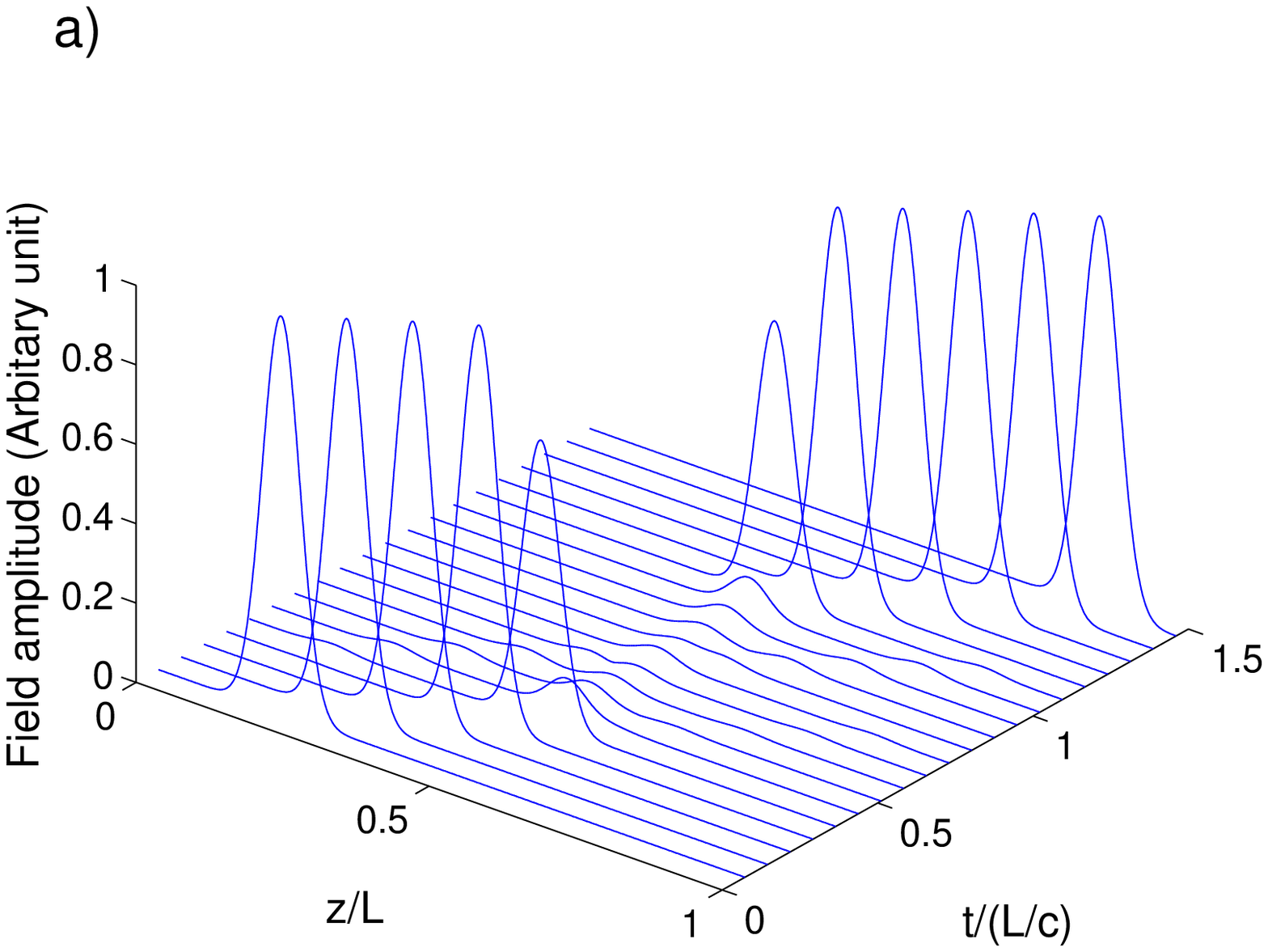}}
\subfigure{\includegraphics[scale=0.4]{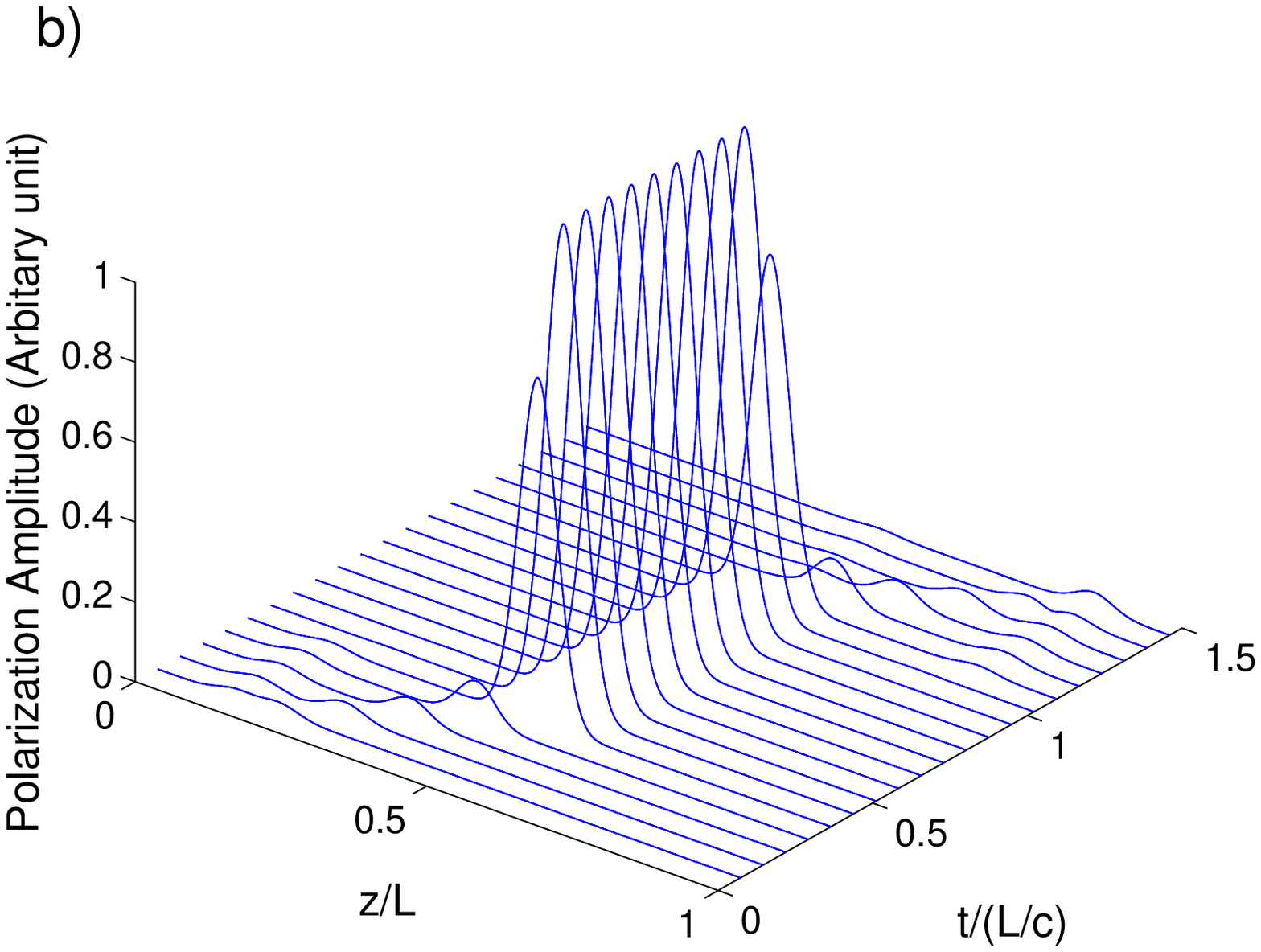}}

\subfigure{\includegraphics[scale=0.3]{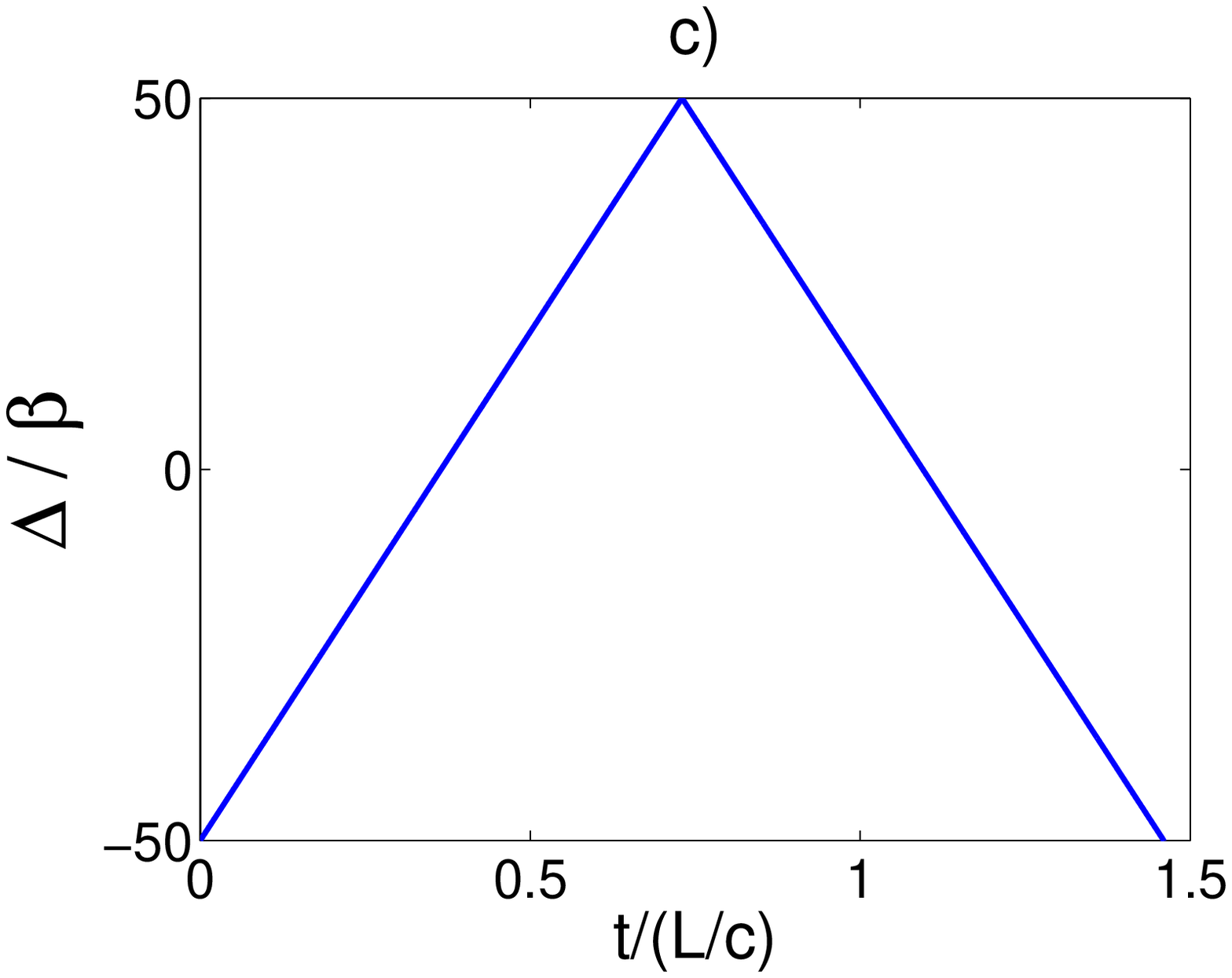}}
\caption{Propagation of field (a) and polarization (b) in the medium in time and space. (c) Detuning as a function of time, We start from $\Delta_{0}=-50\beta$ and sweep to $+\Delta_{0}=50\beta$ with the rate of $\dot{\Delta}=0.4\beta^{2}$. The coupling constant is set to $\beta=30\Delta\omega$. The initial pulse envelope is $\exp[-(z/z_{0})^{2}]$ where $z_{0}$ is $0.045L$. The light is converted into a stationary atomic excitation by sweeping the detuning across the resonance. It is then retrieved by sweeping in reverse direction. }
\label{Polariton_3D}
\end{figure}

We have performed numerical simulations that are in excellent agreement with the described polaritonic picture. Figure \ref{Polariton_3D} is a simulation of the original Maxwell-Bloch equations Eq. (\ref{field_equation}), Eq. (\ref{polarization_equation}). It can be seen that by sweeping the atomic frequency across the resonance, light is slowed down and converted to atomic coherence. By sweeping the atomic frequency in reverse direction, the light is retrieved, as predicted by the polaritonic picture given above.

\begin{figure}[H]
\centering
\includegraphics[scale=0.35]{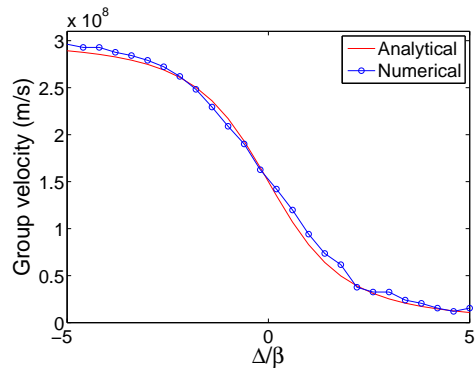}
\caption{Comparison between group velocities resulting from analytical (red) and numerical (blue) calculations. The analytical curve is based on the polaritonic description discussed in the text, while the numerical curve is based on the exact Maxwell-Bloch equations.}
\label{fig:Group_Velocity}
\end{figure}

In addition  we have compared the group velocity obtained from numerical calculations with the group velocity found in the polaritonic picture. Figure \ref{fig:Group_Velocity} shows the excellent agreement between the numerical and analytical group velocities.

Thus far, we have imposed three conditions for obtaining the polaritonic picture. They are summarized below:

\begin{equation}
\beta \ll \Delta_{0},
\label{phase_mismatching}
\end{equation}

\begin{equation}
\dot{\theta} \ll \lambda_{1}, \lambda_{2},
\label{adiabaticity}
\end{equation}

\begin{equation}
\Delta \omega \ll \beta.
\label{disspersion}
\end{equation}

We have investigated the physical significance of these conditions numerically. Figure 5 in the Supplementary Information shows that if we start with an initial detuning that is not sufficiently larger than the coupling, violating condition (\ref{phase_mismatching}), the light couples to both $\Psi$ and $\Phi$ modes, which results in increased loss.

We also examine the adiabatic condition (\ref{adiabaticity}) in Figure 6 in the Supplementary Information. If the adiabatic condition is violated at some time during the storage (retrieval) process, the mode of interest leaks to the other eigenmode, which accelerates to the speed of light (decelerates to the speed of zero), again resulting in loss.

Lastly we examine condition (\ref{disspersion}), which allowed the
expansion of the eigenenergies in k space, in Figure 7 of the Supplementary Information. Violating this condition results in dispersion of the pulse in the medium and thus decreases the fidelity of the storage.

So far we have neglected the decay rate $\gamma$ in our discussion. In the regime where the coupling constant is much larger than the decay rate ($\beta \gg \gamma$), the decay results in decay rates $\gamma\sin^{2}\theta(0)$ and $\gamma\cos^{2}\theta(0)$ for the polaritons $\Psi$ and $\Phi$, without changing the group velocity. The dependence of the decay on the mixing angle $\theta(0)$ is due to the fact that the decay is associated with the atomic (polarization) part of the polaritons.

\section{Experimental Requirements}

In order to store the light efficiently, the pulse has to fit inside the medium, otherwise those frequency components of the pulse that were not in the medium at the time in which the atoms had the corresponding frequency, do not become absorbed in the medium. This imposes a condition on the frequency bandwidth of the input pulses ($\Delta\omega\gg\frac{c}{L}$). This condition along with the requirement for avoiding dispersion of the pulse in the medium ($\beta\gg\Delta\omega$), and the requirement that the coupling rate is significantly greater than the decay rate ($\beta \gg \gamma$), implies a demanding requirement for the optical depth,
\begin{equation}
d=\frac{\beta^{2}L}{c\gamma}\gg \frac{\beta}{\gamma} \gg 1.
\label{optical_depth}
\end{equation}
It is worth mentioning that the same condition on the optical depth can be obtained by using the adiabaticity condition ($\dot{\theta} \ll \lambda$) and the initial detuning requirement ($\Delta_{0} \gg \beta$).

We numerically calculated the efficiency of the AFS protocol as a function of optical depth. The results are shown in Figure 4. This graph shows that high efficiency demands relatively large optical depth.

\begin{figure}[H]
\centering

\subfigure[]{\includegraphics[scale=0.7]{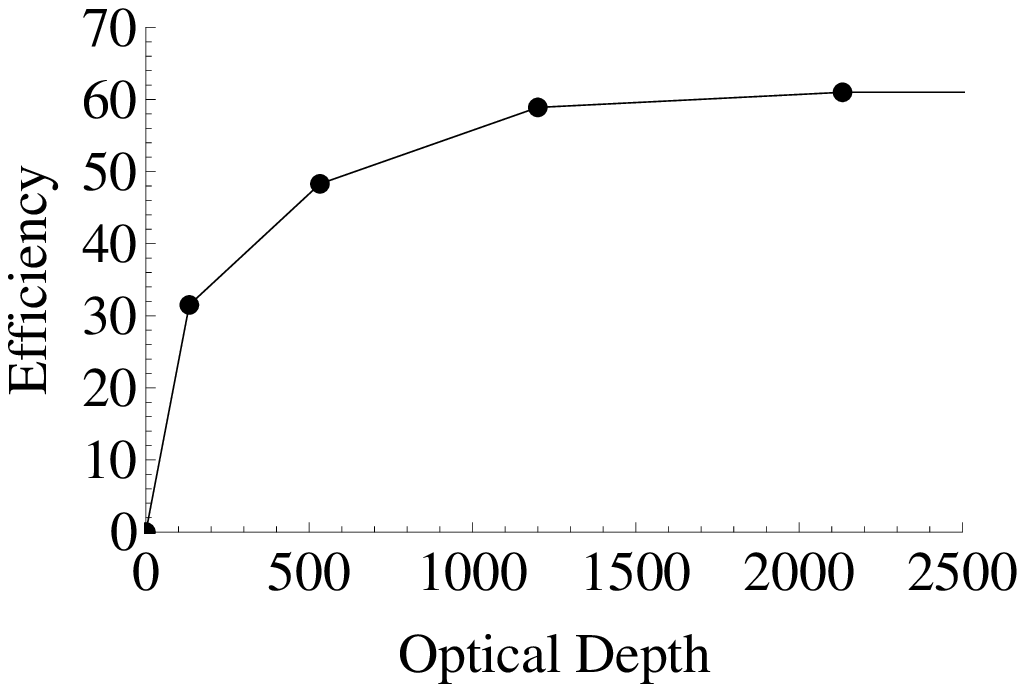}}
\subfigure[]{\includegraphics[scale=0.7]{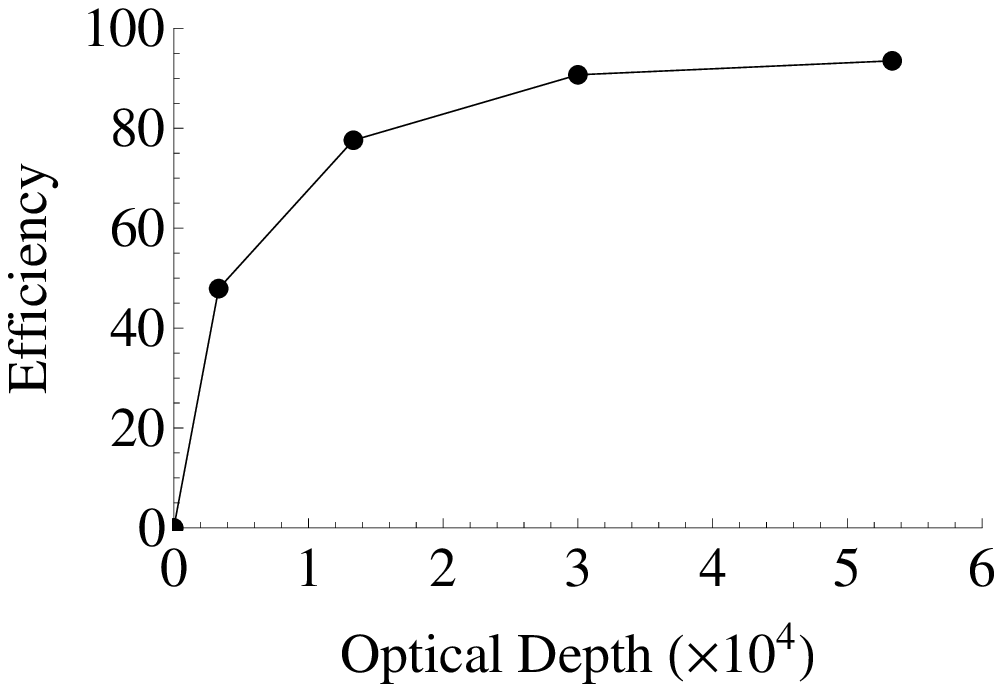}}

\caption{Efficiency as a function of optical depth for (a) decay rate of $\gamma=5 \times 10^{-2} \Delta \omega$ and (b) decay rate of $\gamma=2 \times 10^{-3} \Delta \omega$. Here we start from initial detuning  $\Delta_{0}=-8\beta$ and sweep to $+\Delta_{0}=8\beta$ in a time of $2L/3c$ (For the retrieval process we do the reverse). We fixed the length to 20 cm in accordance with Ref.\cite{Photonic_crystal} and changed the optical depth by changing the coupling constant. The initial envelope is $\exp[-(z/z_{0})^{2}]$, where $z_{0}$ is $0.15 L$.}
\end{figure}

Intuitively the high optical depth requirement can be understood from the fact that the conditions (\ref{phase_mismatching}-\ref{disspersion}) impose a large memory bandwidth (swept frequency range) compared to the pulse spectral bandwidth. Large optical depth is necessary to compensate for the fact that the atoms are ``spread out'' over this large frequency range. However, there is room for lowering the precise optical depth requirements by optimizing the sweeping of the atomic frequency in time. This is a subject for future studies. The saturation value of the efficiency in Figure 4 depends on the decay rate as well as the initial detuning. As we mentioned earlier, the initial detuning determines the portion of the input light that is coupled to the polariton mode of interest.

One promising candidate system for implementation of the AFS protocol are
atoms in a hollow-core photonic crystal fiber (HCPCF) \cite{Photonic_crystal,lukinprl,lukinpra}. On the one hand, this system is suitable for accommodating the entire input pulse, as the length of the HCPCF can be extended to adequate amounts. On the other hand relatively large optical depth (up to $300$) has already been demonstrated in such systems \cite{{Photonic_crystal}}. For similar length and optical depth as in reference \cite{{Photonic_crystal}}, an efficiency of $\approx 40 \% $ (with a fidelity of $\approx 99 \%$) is achievable.

Another challenge for the implementation of the AFS memory protocol is the large range over which the atomic frequency has to be swept. As we discussed earlier, to avoid mixing the eigenmodes of the system, the range of swept atomic frequency should be larger than the coupling constant ($\Delta_{0}\gg\beta \gg \Delta \omega$). For the system proposed here (Ref.\cite{Photonic_crystal}), the swept frequency range should be of the order of 100 GHz.  This challenge can be overcome by increasing the length of medium, which allows decreasing the coupling constant without compromising the optical depth.

\section{Conclusions and Outlook}

The polaritonic description given above made it clear that the AFS memory has significant similarities with EIT based memories. As we mentioned in the introduction, the AFS memory is also connected to the GEM memory in the regime in which the size of the pulses is smaller than the length of the medium. Numerical calculations (shown in Figure 8 of the Supplementary Information) also verify this connection. This suggests the notion of a real-space polariton in the GEM memory which differs from the momentum-space polariton introduced in Ref. \cite{Polariton_GEM}. Nevertheless there are some differences between AFS and GEM protocols even in the small pulse regime. In AFS the pulse slows down and stops when the atoms are far detuned from the pulse (after having been swept through the resonance). In contrast, in GEM the pulse stops close to the resonant part of the medium, or, if the optical depth is high enough, even before the resonant part of the medium is reached. This fact leads to the shift in the release time of the output that can be seen in Figure 8 of the Supplementary Information.

The AFS memory is linked to photon echo based memories more generally. When we sweep the atomic frequency in time, the atomic coherences become dephased and thus re-emission is inhibited. By sweeping in reverse direction, the atomic coherences become rephased and the light is re-emitted. This fact imposes some limits on sweeping the atomic frequency. The sweep should be fast enough compared to the coupling constant to ensure that dephasing happens before reemission. The details of the outlined dephasing-rephasing process for the AFS memory are a subject for future studies.

Our protocol also shares some similarities with the proposal of Ref. \cite{Fan}, where light guided by Coupled Resonator Optical Waveguides (CROW) is stored by dynamically changing the resonance frequency of side cavities that are coupled to the CROW. In the AFS protocol the atoms play the role of the side cavities in Ref. \cite{Fan}.

It is also worth pointing out that there are some similarities between the AFS memory and Ref. \cite{Thierry}, which proposes to store light by dynamically controlling the splitting of two atomic transitions. These authors also develop a polaritonic picture for their protocol.

In conclusion we have proposed and analyzed a quantum memory protocol based on sweeping the resonance frequency of two-level atoms. We have shown that this new AFS memory protocol shares features with several other existing memory protocols, including EIT and photon-echo based memories. Besides being interesting in its own right, this new protocol can thus also be seen as a step towards a unified description of the complex zoology of quantum memory protocols.

\section*{Acknowledgments} We thank M. Hedges for a useful discussion. This work was supported by NSERC and AITF.

\vspace{1cm}

\section*{Supplementary Information}

As discussed in the paper, for obtaining the polaritonic picture we have assumed the following three conditions:

\begin{equation}
\beta \ll \Delta_{0},
\label{phase_mismatching_1}
\end{equation}

\begin{equation}
\dot{\theta} \ll \lambda_{1}, \lambda_{2},
\label{adiabaticity_1}
\end{equation}

\begin{equation}
\Delta \omega \ll \beta.
\label{disspersion_1}
\end{equation}

Condition (\ref{phase_mismatching_1}) is essential for initially exciting only one of the polaritons. Condition \ref {adiabaticity_1} guarantees the adiabaticity of the process, thus avoiding leakage from one polariton to the other polariton. Condition \ref{disspersion_1} prevents dispersion, which is important for achieving high fidelity. Here we numerically study these conditions.

\begin{figure}[H]
\centering
\subfigure[]{\includegraphics[scale=0.3]{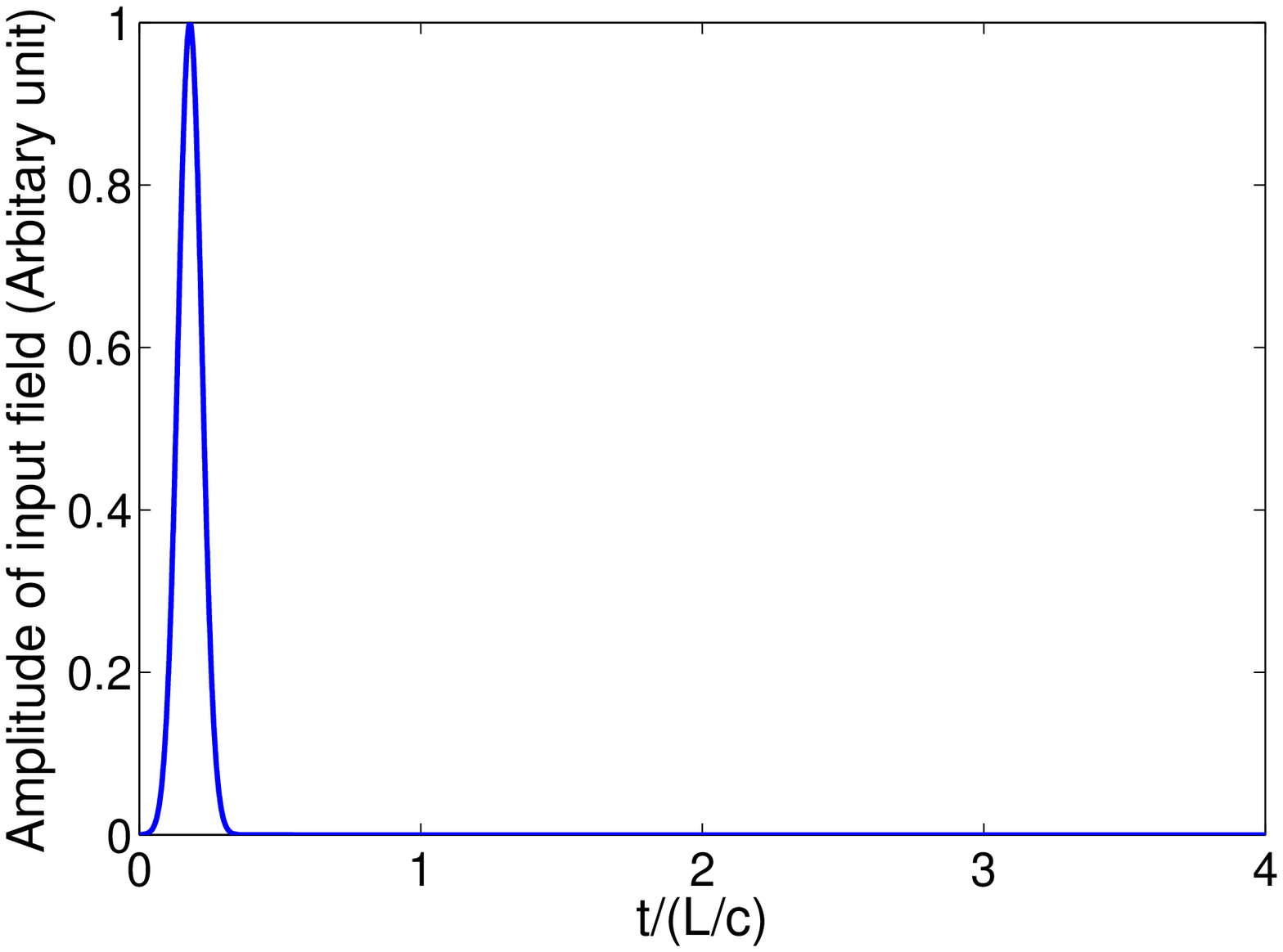}
\label{fig:mismatching_a}}

\subfigure[]{\includegraphics[scale=0.3]{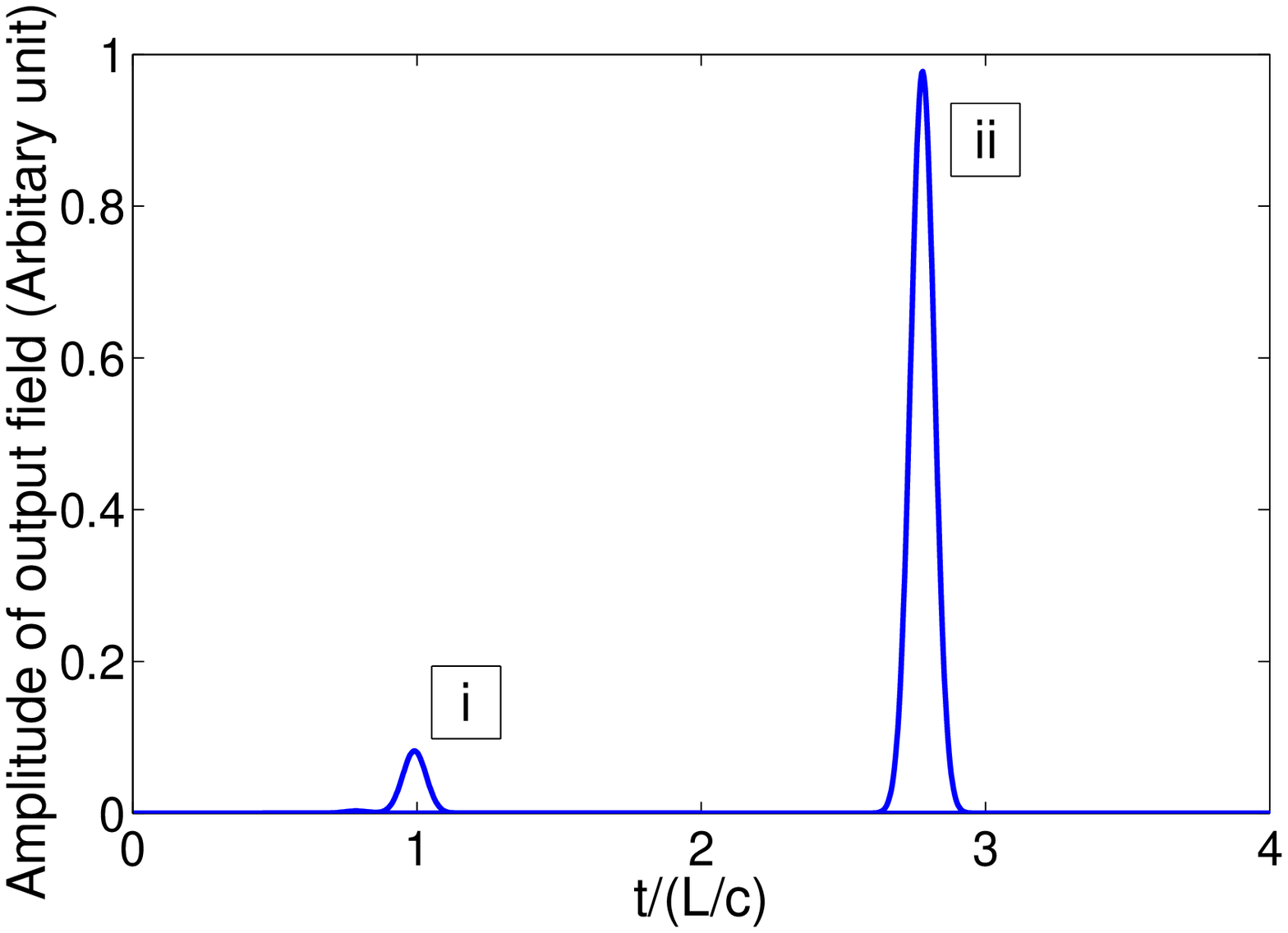}
\label{fig:mismatching_b}}
\subfigure[]{\includegraphics[scale=0.3]{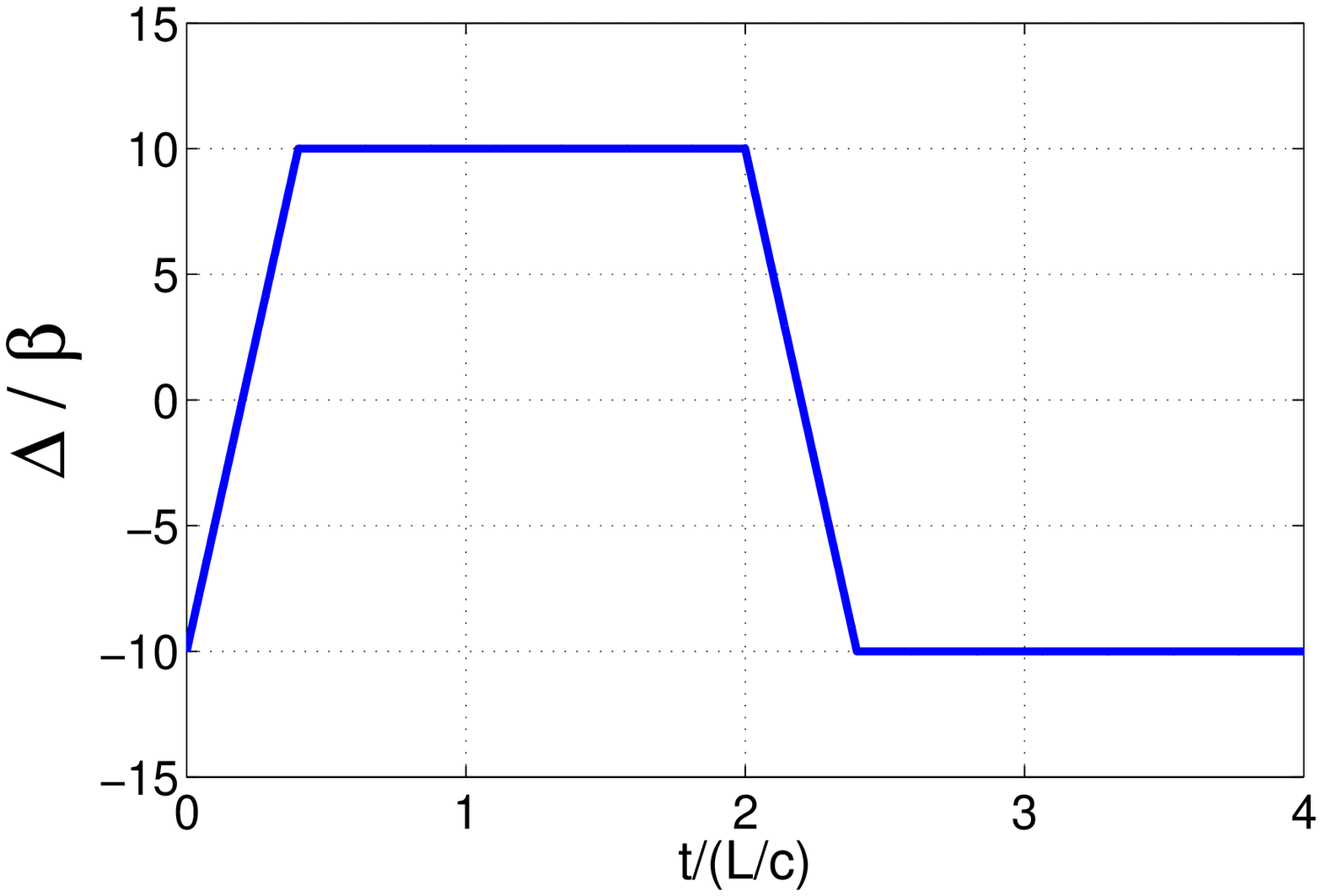}
\label{fig:mismatching_c}}

\subfigure[]{\includegraphics[scale=0.3]{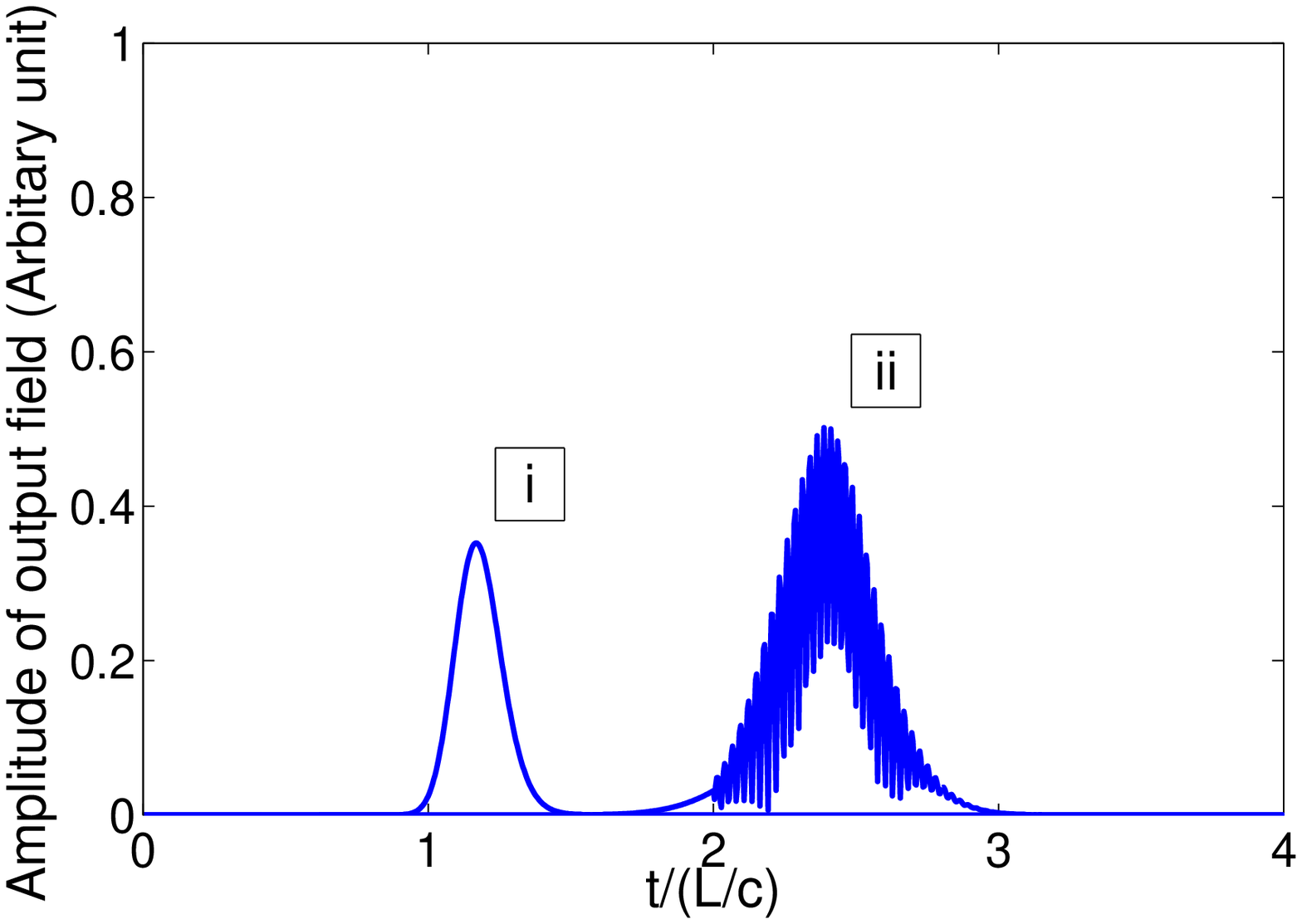}
\label{fig:mismatching_d}}
\subfigure[]{\includegraphics[scale=0.3]{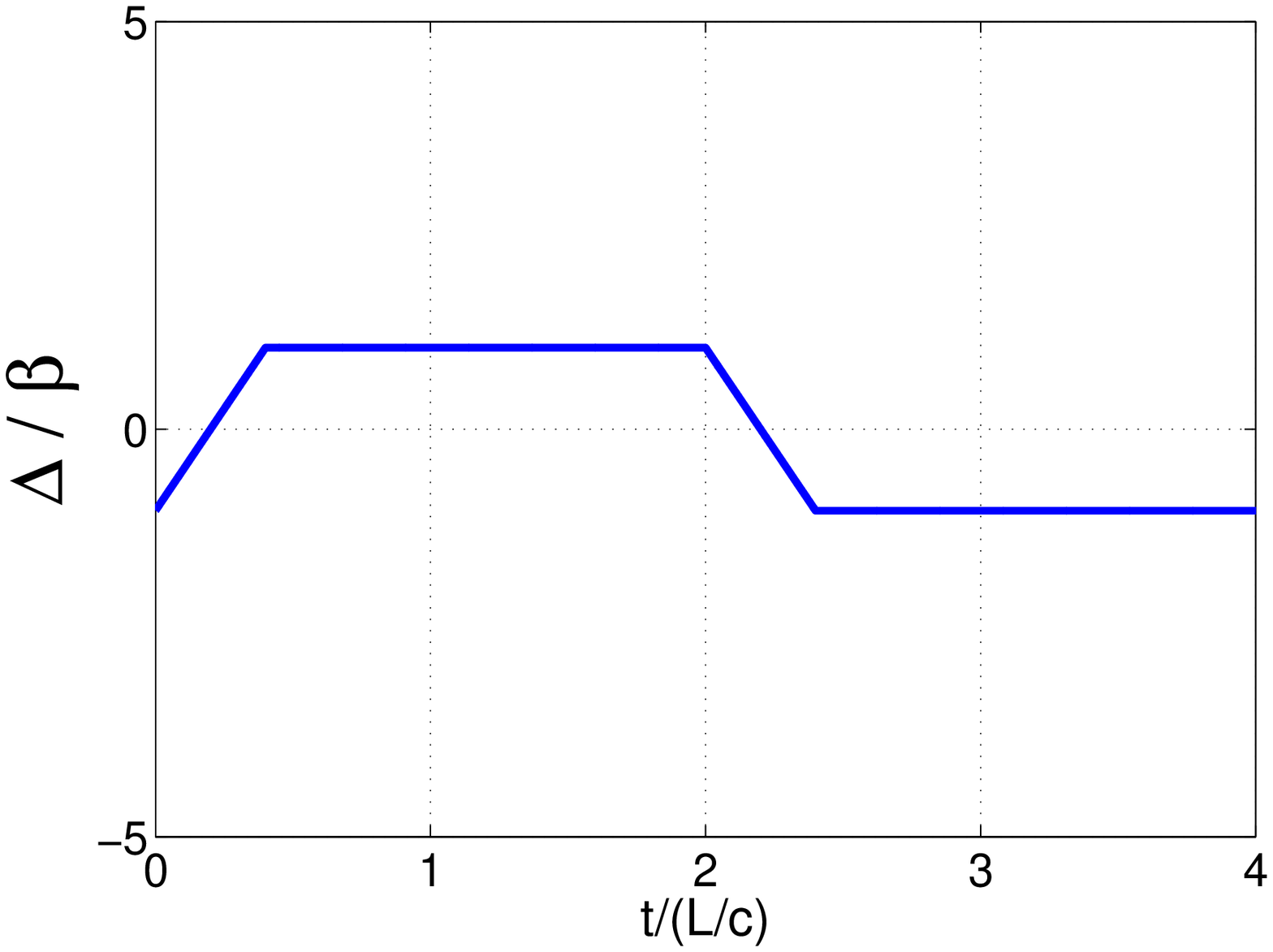}
\label{fig:mismatching_e}}

\subfigure[]{\includegraphics[scale=0.3]{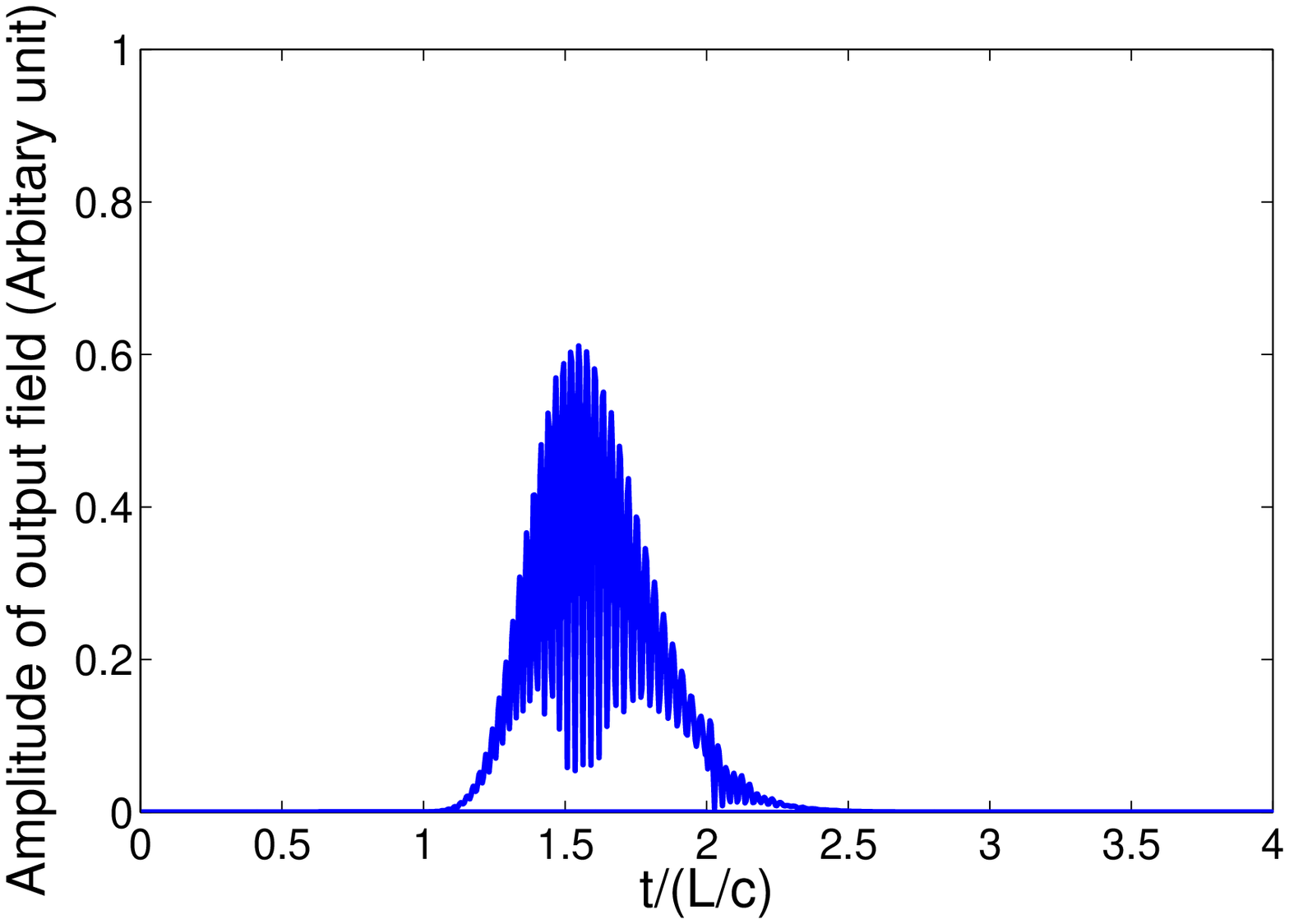}
\label{fig:mismatching_f}}
\subfigure[]{\includegraphics[scale=0.3]{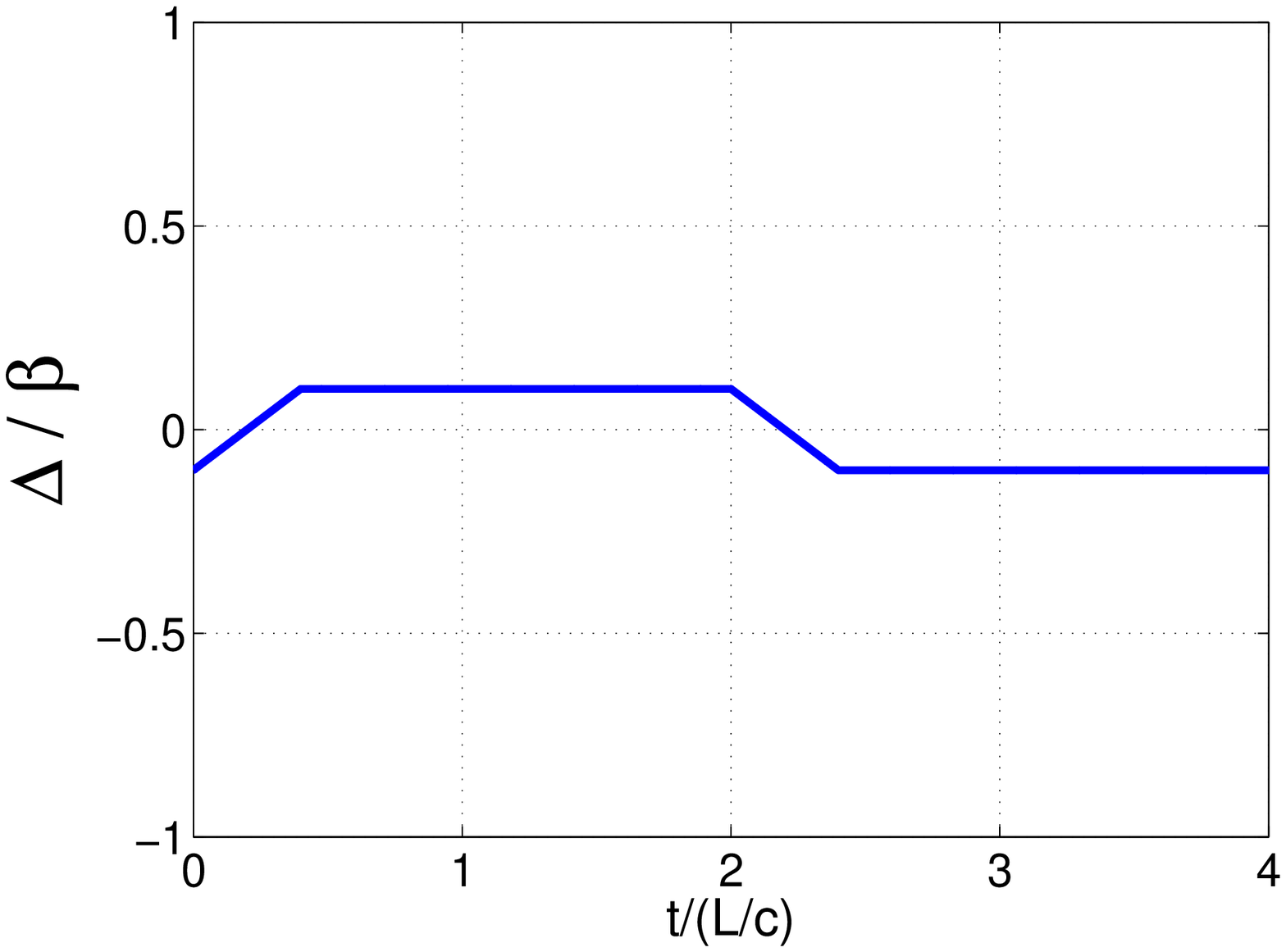}
\label{fig:mismatching_g}}

\begin{singlespace}
\small
\caption{Effect of initial detuning on the output of the memory. The bandwidth of the input field is set to $\Delta \omega =0.1\beta$. (a) shows the temporal shape of the input pulse. Note that the input pulse is initially in the medium . We show the output field for different values of initial detuning b)$\Delta_{0}=-10\beta$, d)$\Delta_{0}=-\beta$, f)$\Delta_{0}=-0.1\beta$. (c), (e), (g) show how the detuning changes in each case. (i) is the transmitted and (ii) is the retrieved pulse.}
\end{singlespace}
\end{figure}

First, figures \ref{fig:mismatching_a}-\ref{fig:mismatching_g} show what happens when condition (\ref{phase_mismatching_1}) is violated. It can be seen that by decreasing the initial detuning (while the other conditions are fulfilled), we excite polariton $\Phi$ (i) more and polariton $\Psi$ (ii) less. According to Eq. (10) and Eq. (11) in the paper, as we change the detuning, polariton $\Psi$ slows down and converts to polarization, while polariton $\Phi$  accelerates from zero group velocity to the speed of light and leaves the medium. Therefore excitation of polariton $\Phi$ leads to loss. When the starting detuning is  much less than the coupling rate (Figure \ref{fig:mismatching_f}), $\theta$ is almost $\pi/4$, so according to Eq. 10 and Eq. 11 both polaritons travel with speed $c/2$ and, since they have opposite phase ($\lambda_{1}=\beta$ and $\lambda_{2}=-\beta$), there is beating between them. This means that the light undergoes a series of re-emission and absorption processes with rate $\beta$ (see also Figure \ref{fig:mismatching_d}).

\begin{figure}[H]
\centering

\subfigure[]{\includegraphics[scale=0.3]{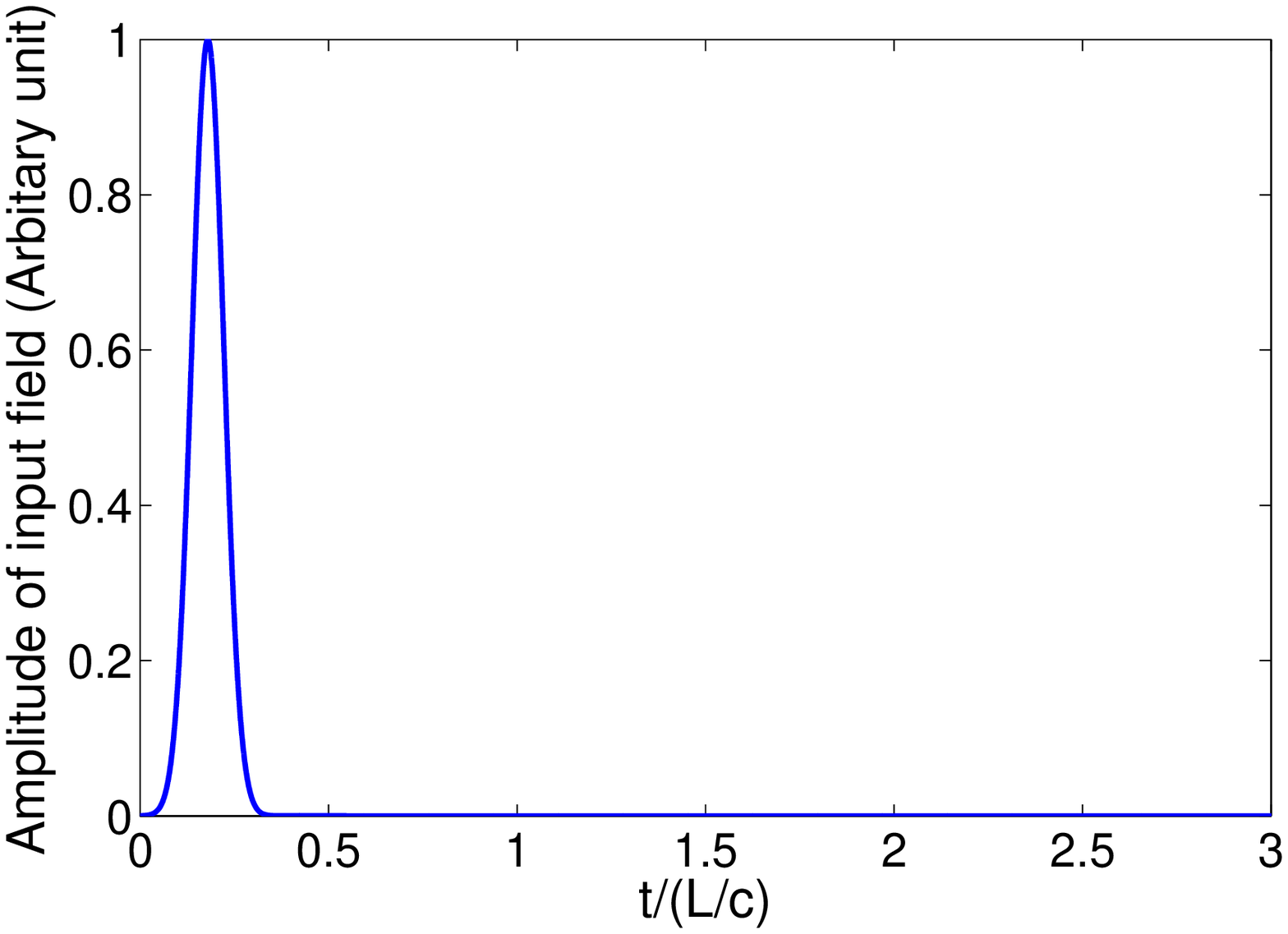}
\label{fig:adiabaticity1}}

\subfigure[]{\includegraphics[scale=0.27]{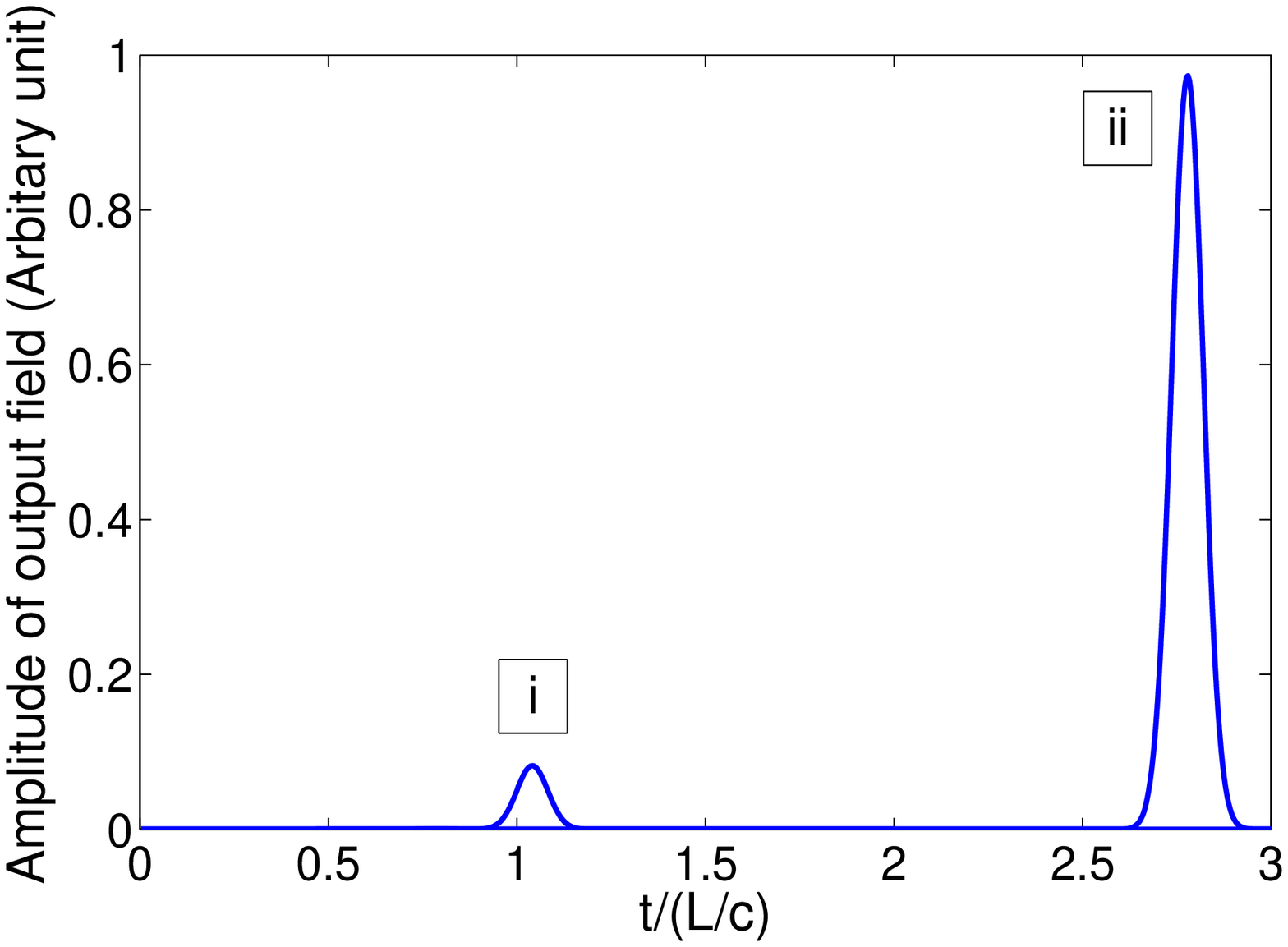}}
\subfigure[]{\includegraphics[scale=0.27]{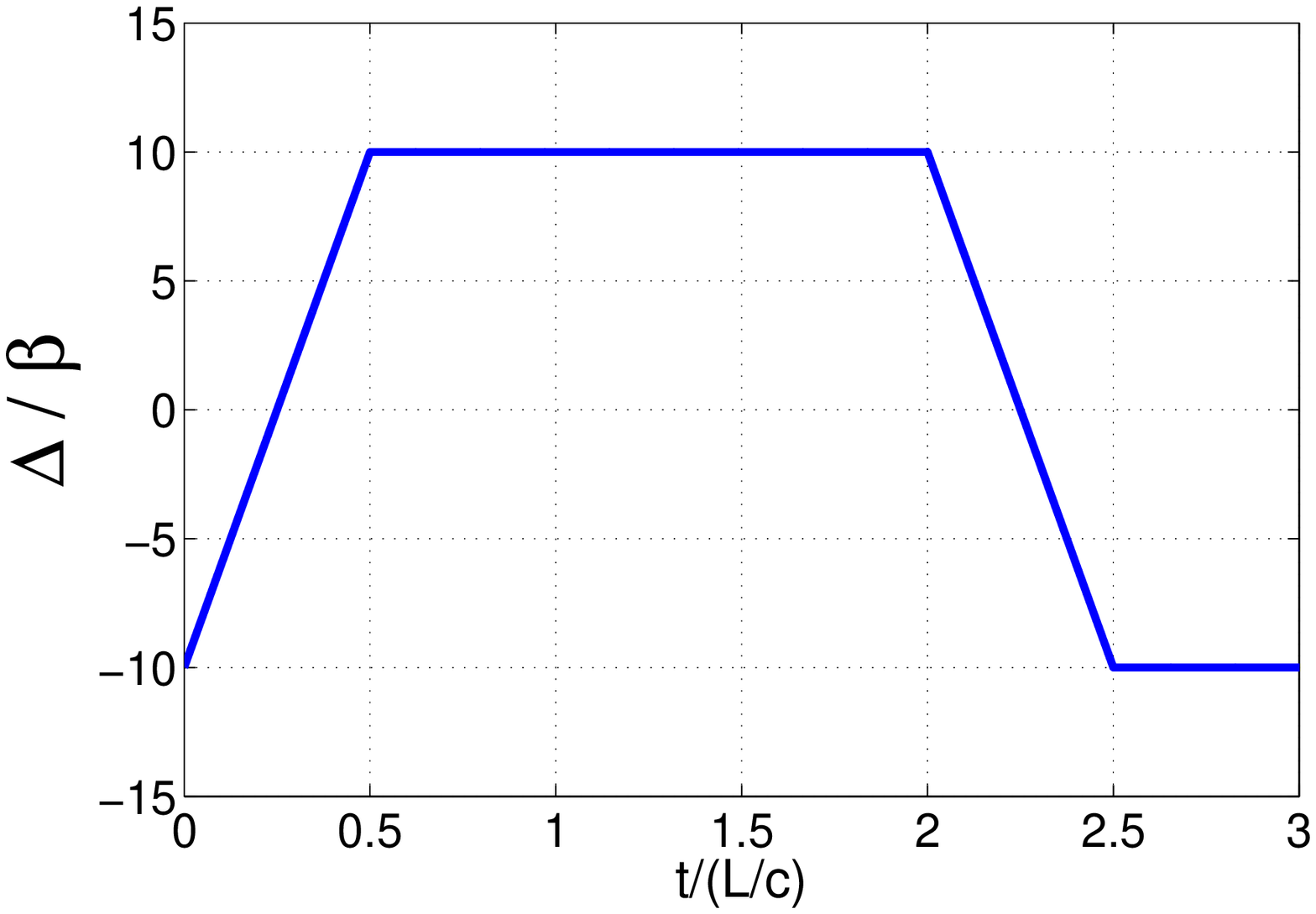}}

\subfigure[]{\includegraphics[scale=0.27]{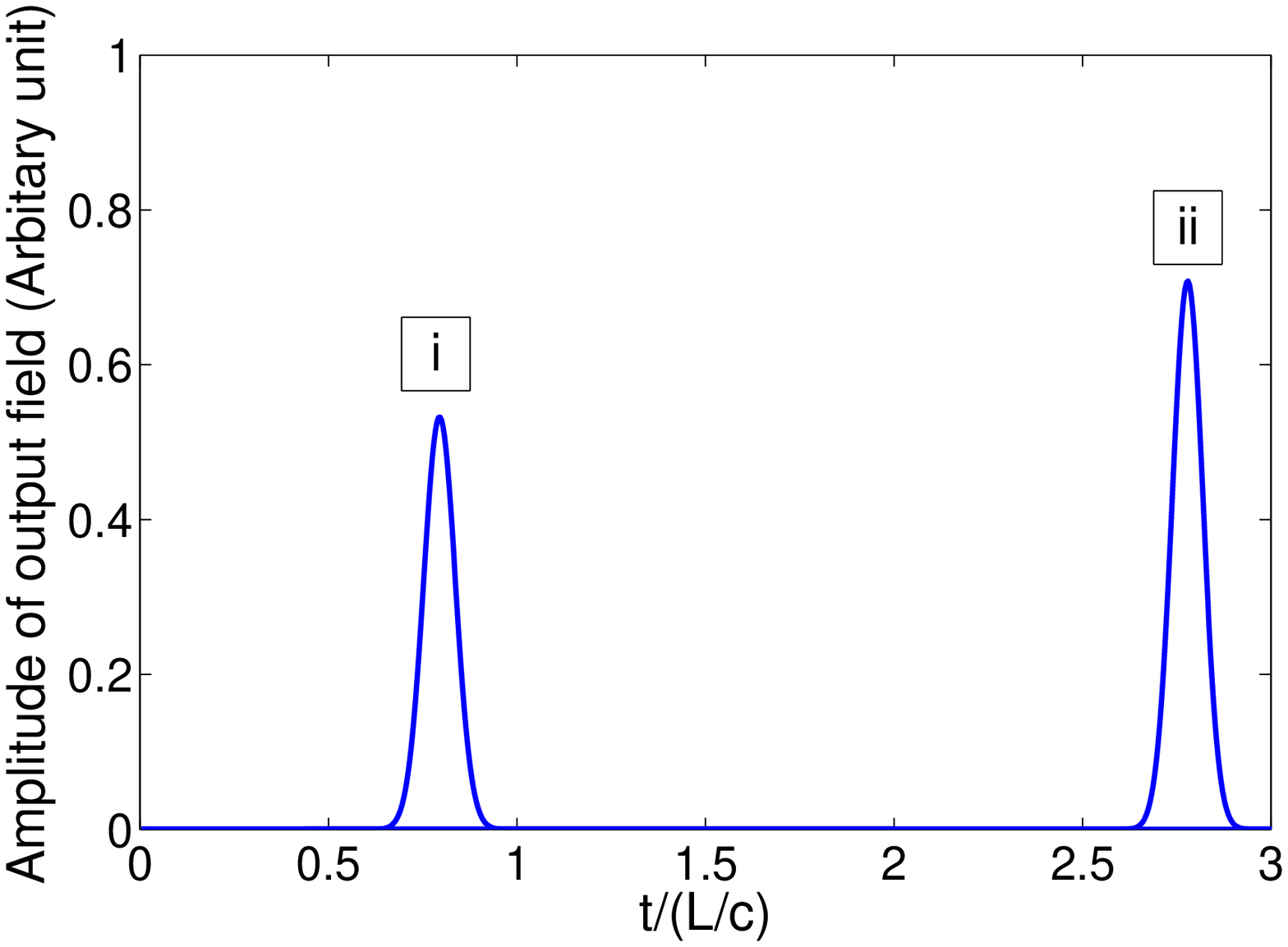}}
\subfigure[]{\includegraphics[scale=0.27]{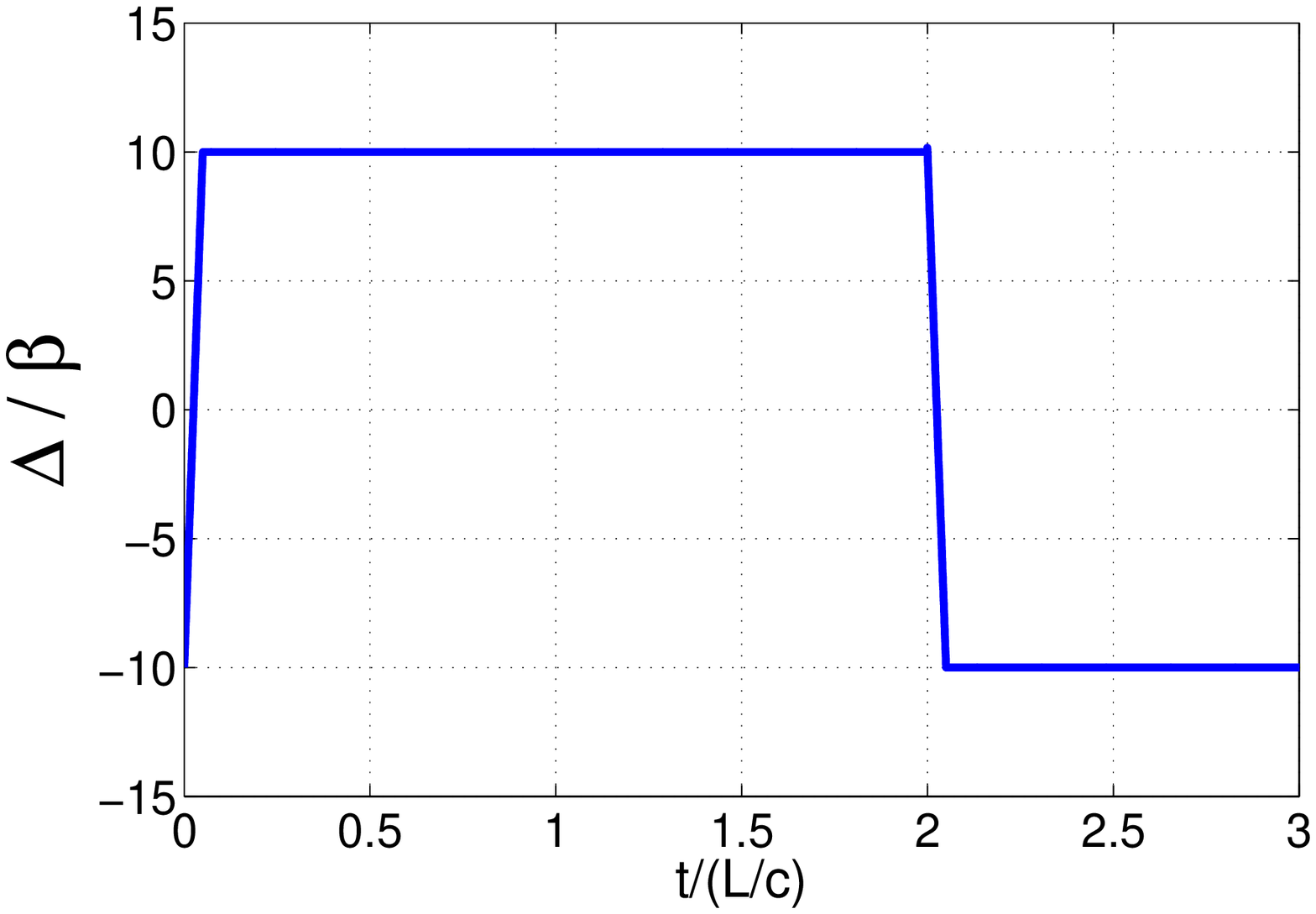}}

\subfigure[]{\includegraphics[scale=0.27]{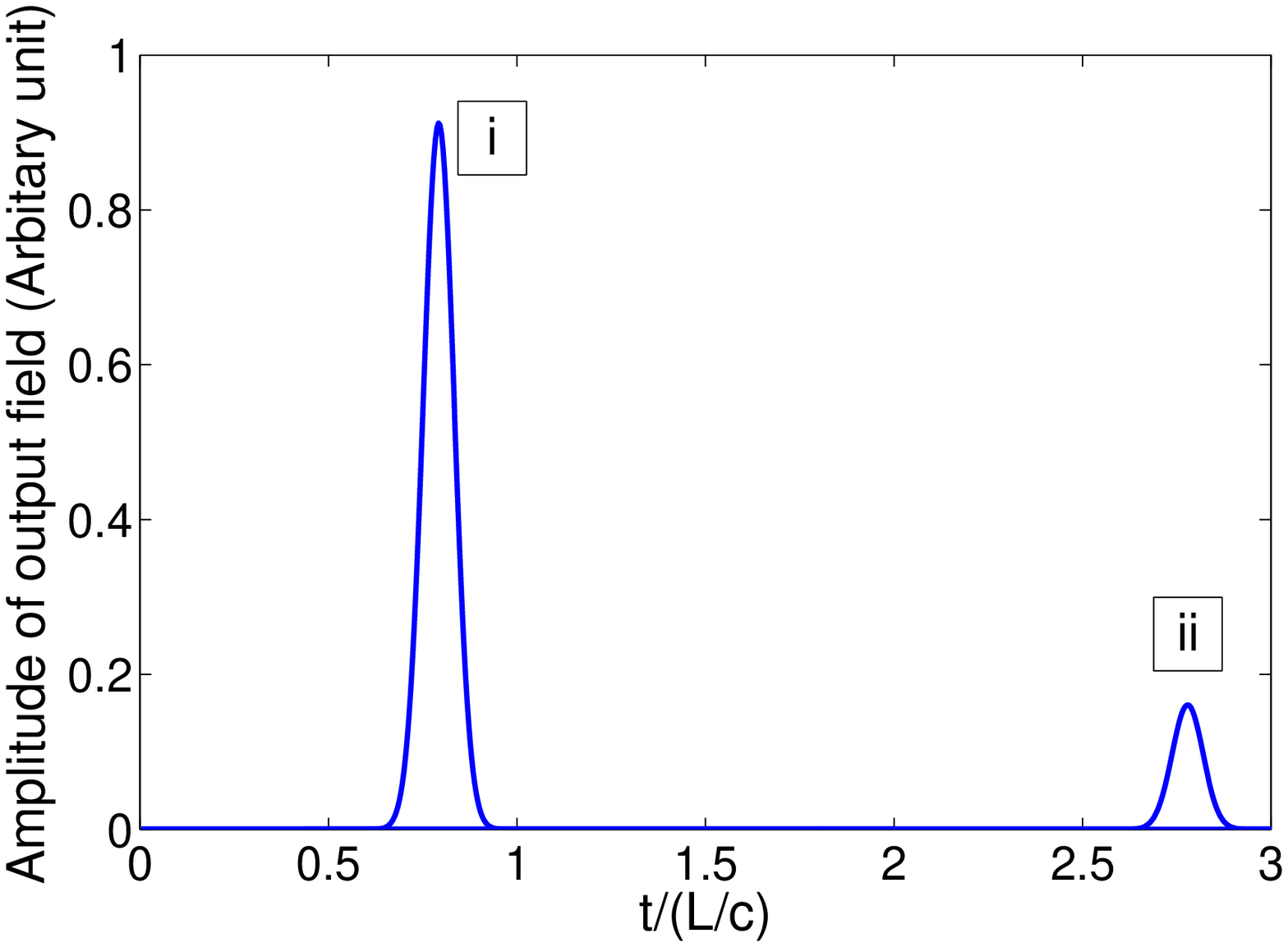}}
\subfigure[]{\includegraphics[scale=0.27]{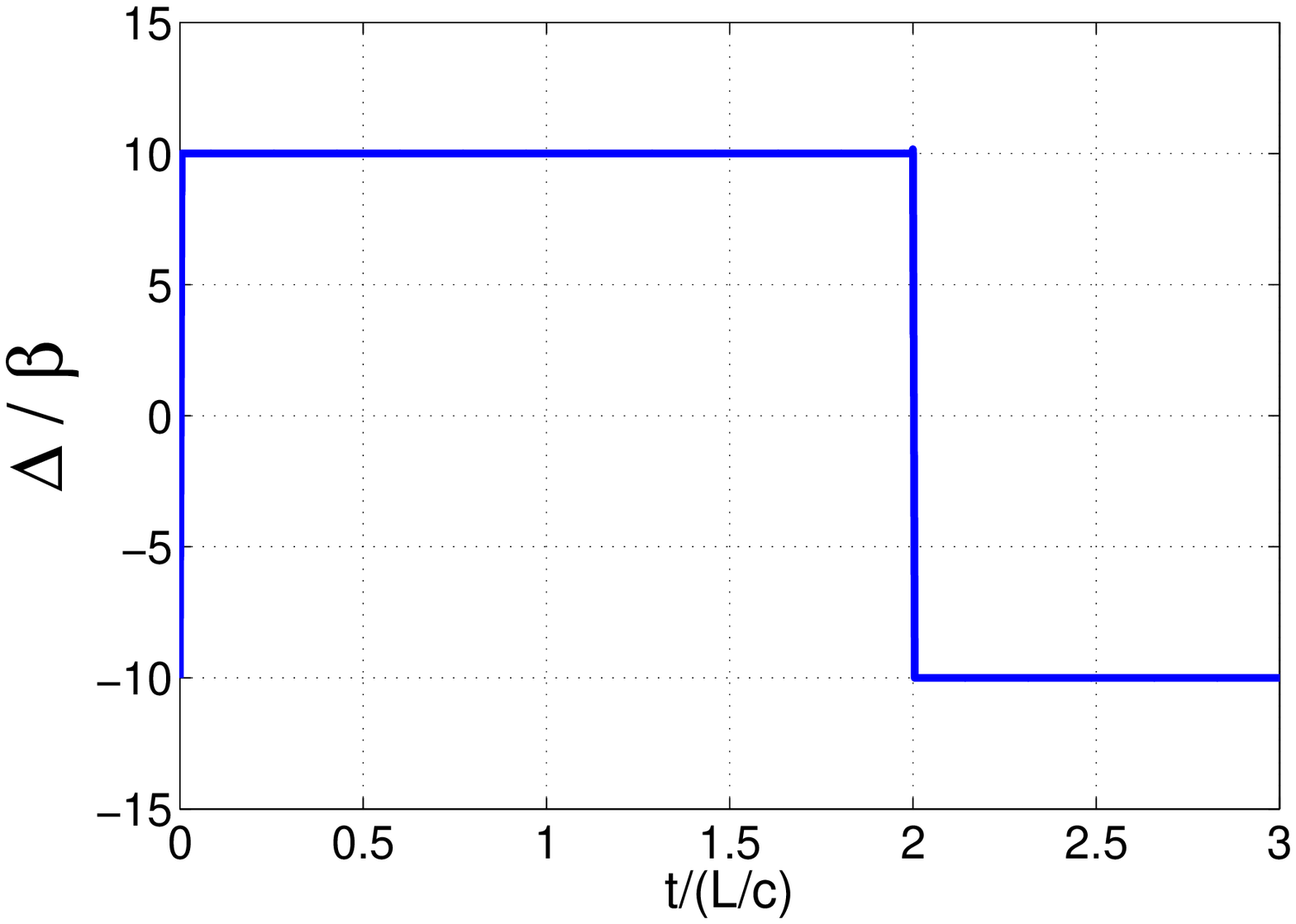}
\label{fig:adiabaticity2}}

\begin{singlespace}
\footnotesize {
\caption{Effect of the rate of changing the detuning on the output of the memory. The bandwidth of the input field is set to $\Delta \omega =0.1\beta$. (a) shows the temporal shape of the input pulse. Note that the input pulse is initially in the medium. The initial detuning for all of the cases is $\Delta_{0}=-10\beta$. We sketch the output field for different values of $\dot{\Delta}$,  b)$\dot{\Delta}=0.3\beta^{2}$, d)$\dot{\Delta}=3\beta^{2}$, f)$\dot{\Delta}=30\beta^{2}$. (c), (e), (g) show the detuning as a function of time. (i) is the transmitted and (ii) is the retrieved pulse.}}

\end{singlespace}

\end{figure}

Secondly we examine the adiabaticity condition \ref{adiabaticity_1}. Figures \ref{fig:adiabaticity1}-\ref{fig:adiabaticity2} show how the output field changes when we violate the adiabaticity condition.

As we increase $\dot{\Delta}$ (while the other two conditions hold), the process becomes less adiabatic, resulting in leakage of polariton $\Psi$ (ii) into polariton $\Phi$ (i), which accelerates to the speed of light and leaves the medium. Therefore this condition is necessary for efficient absorption of the light.

\begin{figure}[H]

\centering
\subfigure[]{\includegraphics[scale=0.3]{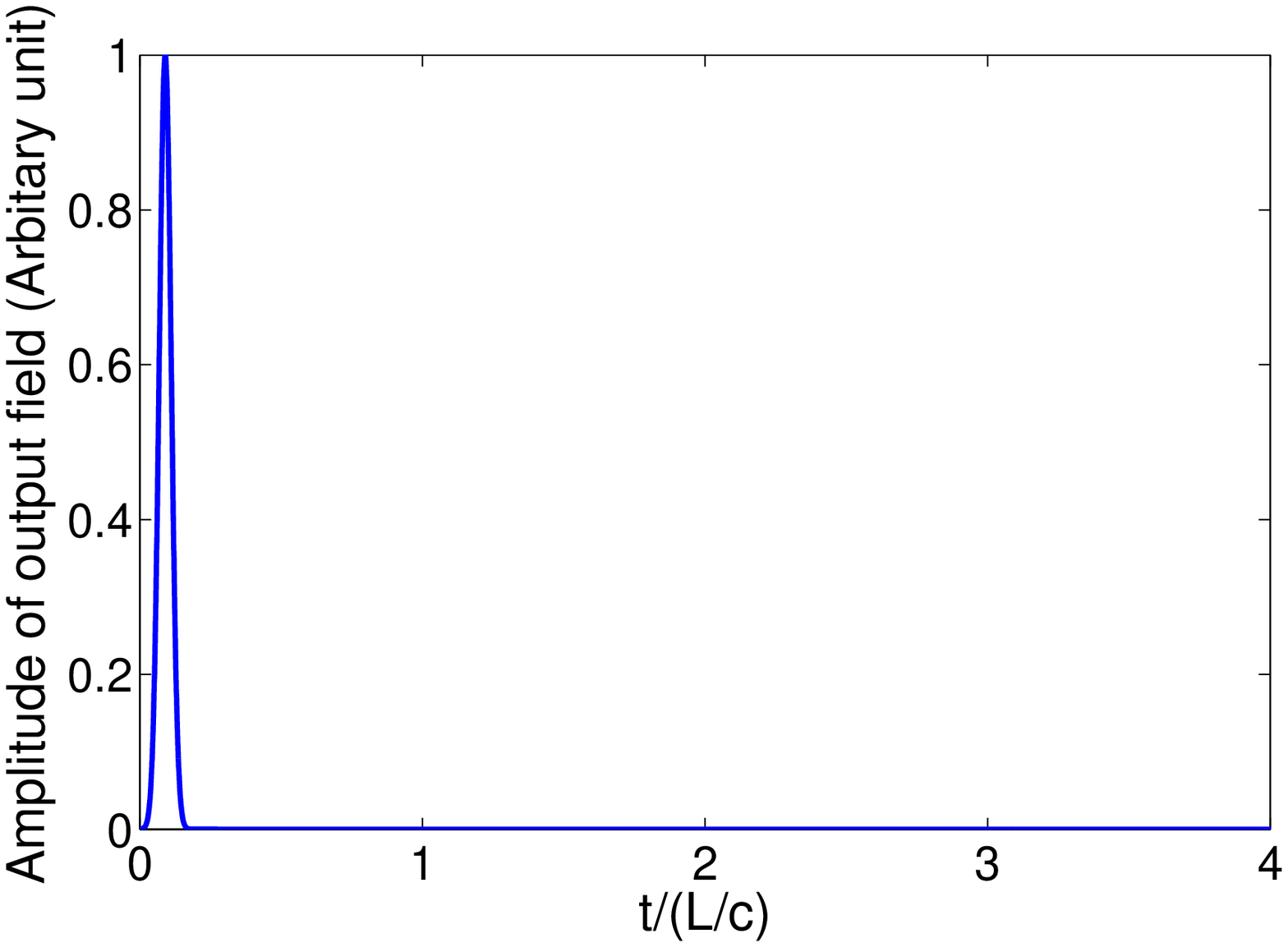}
\label{fig:disspersion1}}

\subfigure[]{\includegraphics[scale=0.3]{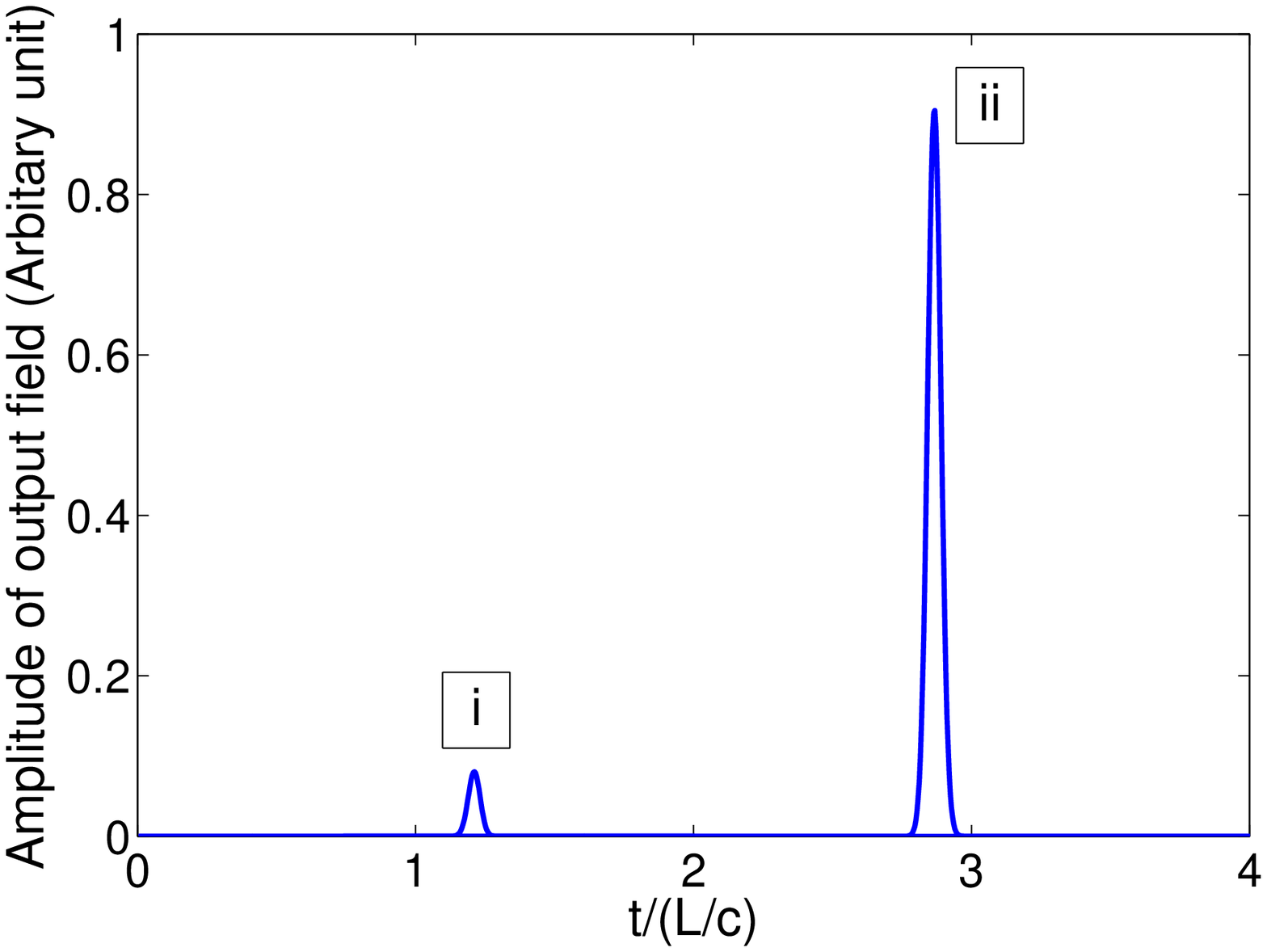}}
\subfigure[]{\includegraphics[scale=0.3]{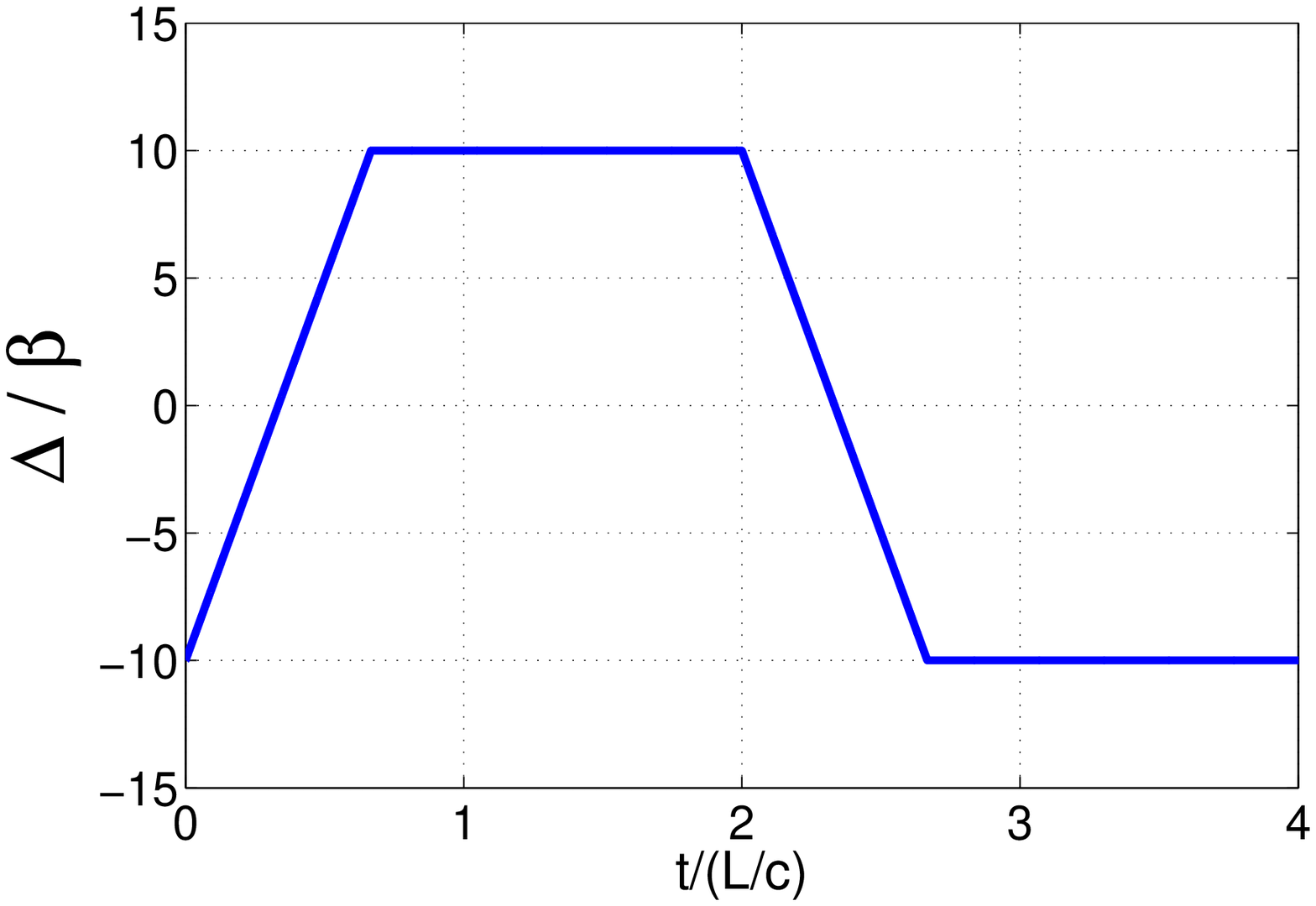}}

\subfigure[]{\includegraphics[scale=0.3]{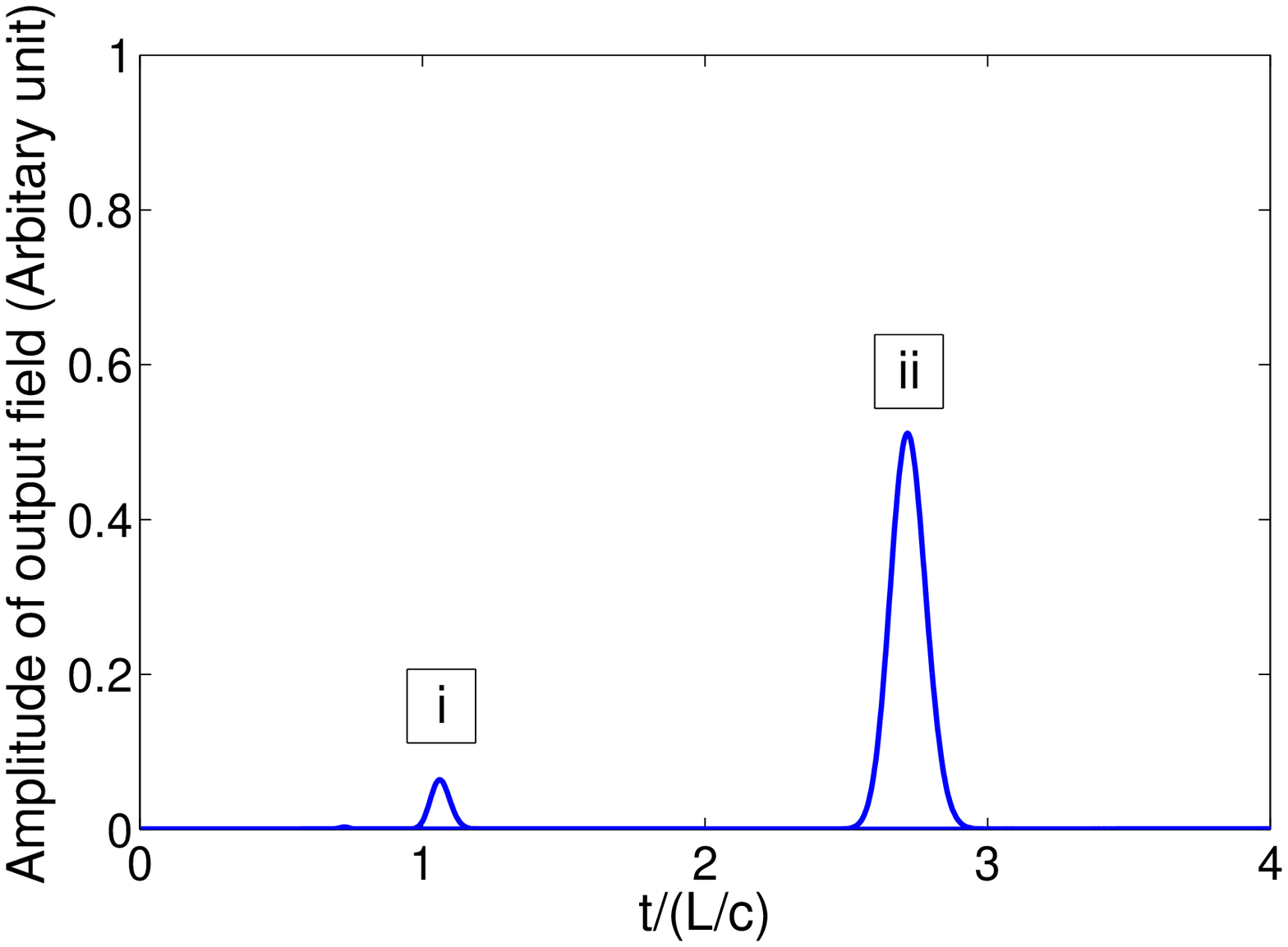}}
\subfigure[]{\includegraphics[scale=0.3]{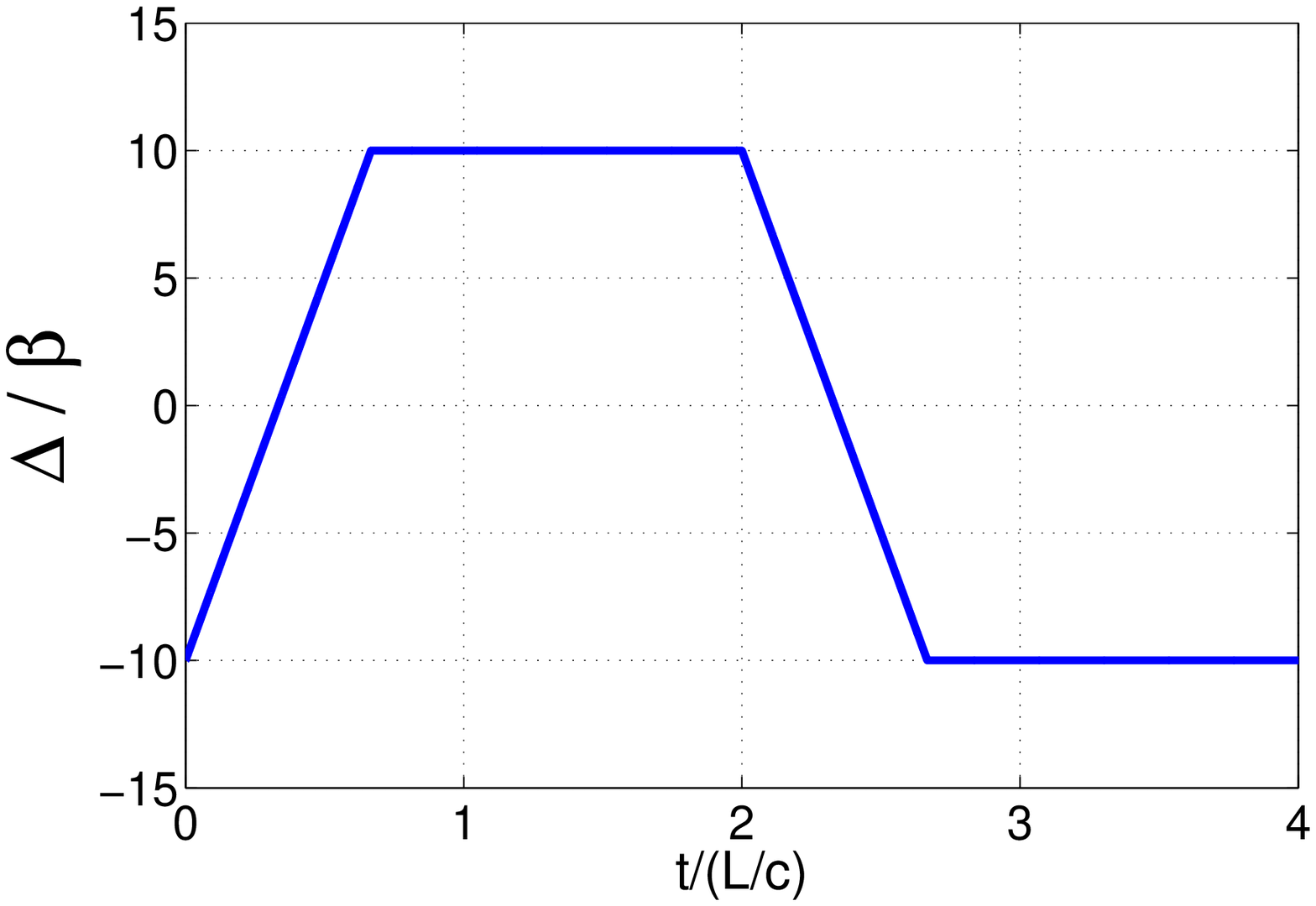}
\label{fig:disspersion2}}

\begin{singlespace}
\footnotesize{
\caption{Effect of the value of $\beta$ (compared to the bandwidth of the input pulse) on the output of the memory. (a) shows the temporal shape of the input pulse. Note that the input pulse is initially in the medium. The initial detuning for all of the cases is $\Delta_{0}=-10\beta$. We sketch the output field for two different values of $\beta$  b) $\beta=16 \Delta \omega$, d)$\beta=4\Delta \omega$. (c), (e) show the detuning as a function of time. (i) is the transmitted and (ii) is the retrieved pulse. }}
\end{singlespace}

\end{figure}

Thirdly we study the effect of violating condition (\ref{disspersion_1}). Figures.\ref{fig:disspersion1}-\ref{fig:disspersion2} show how the output changes when the coupling rate approaches the bandwidth of the pulse.

It can be seen that when $\beta$ approaches the bandwidth of the pulse, the output pulse becomes broader, which reduces the fidelity of the memory protocol. This can be explained by considering the expansion of the eigenvalues in terms of $k$. When $\beta$ is comparable to the bandwidth of the pulse, we no longer can neglect second and higher orders of $k$ in expansion of eigenenergies, resulting in dispersion of the pulse.

\begin{figure}[H]
\centering
\subfigure[]{\includegraphics[scale=0.4]{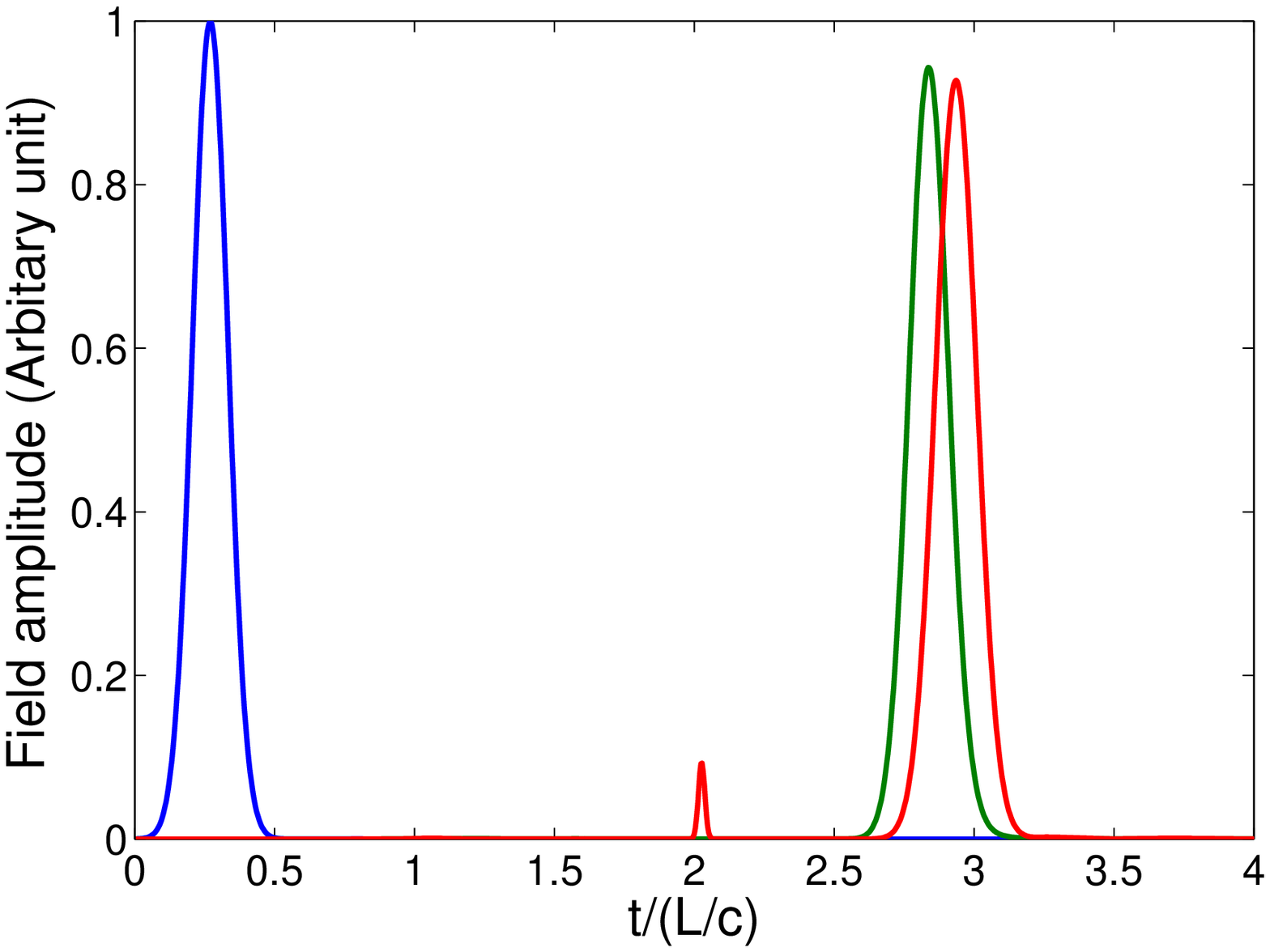}}

\subfigure[]{\includegraphics[scale=0.3]{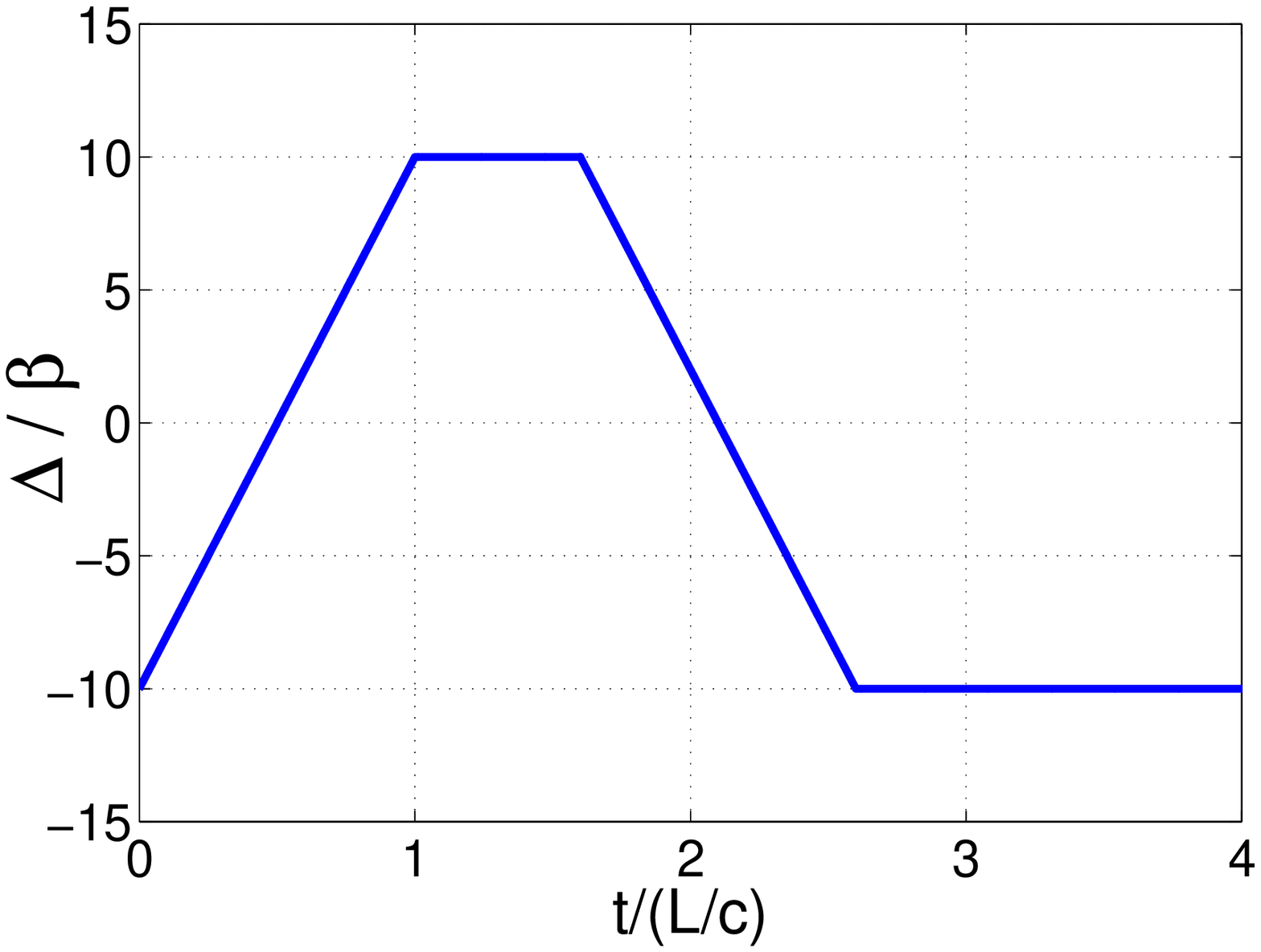}}

\subfigure[]{\includegraphics[scale=0.3]{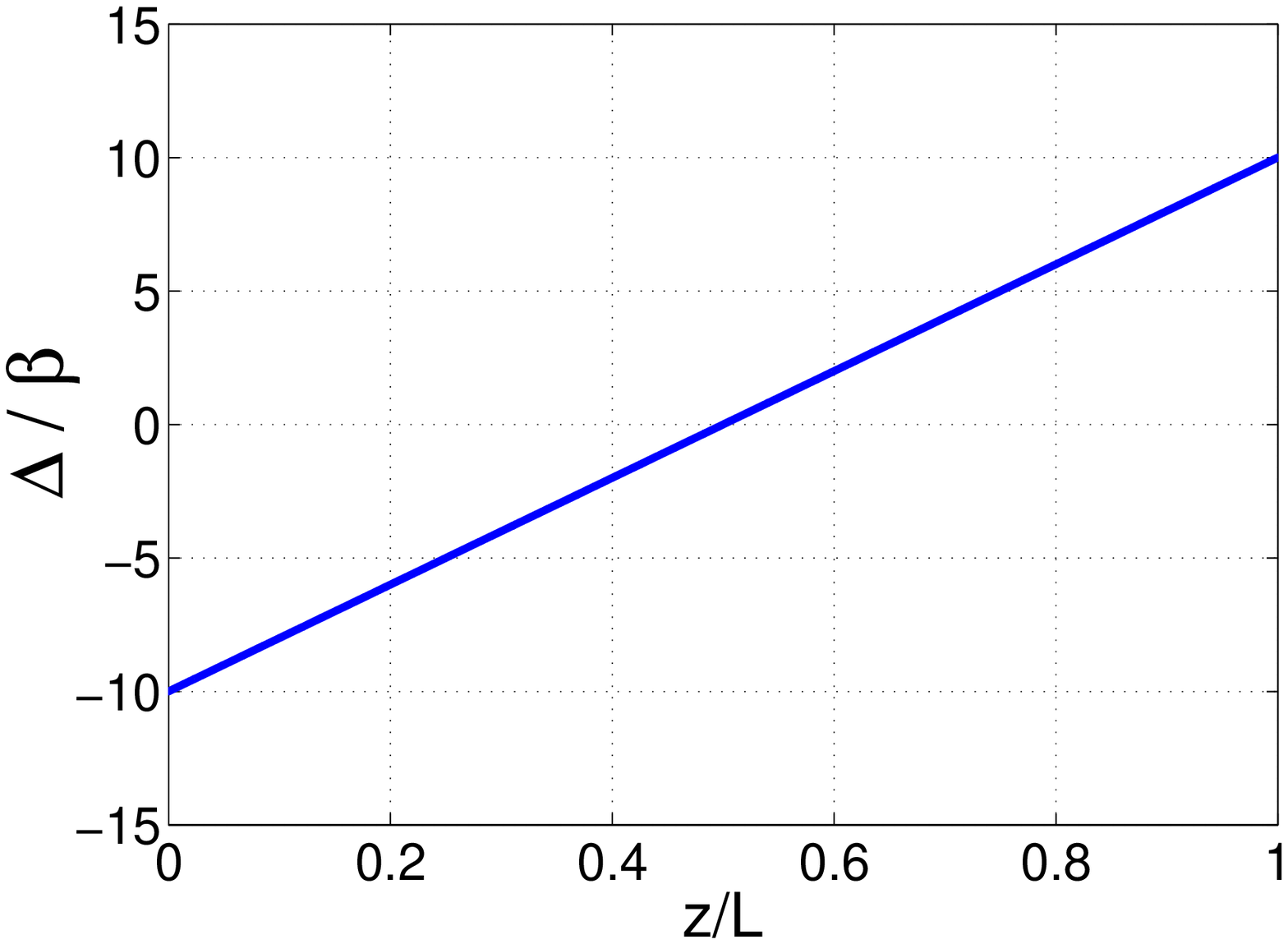}}
\subfigure[]{\includegraphics[scale=0.3]{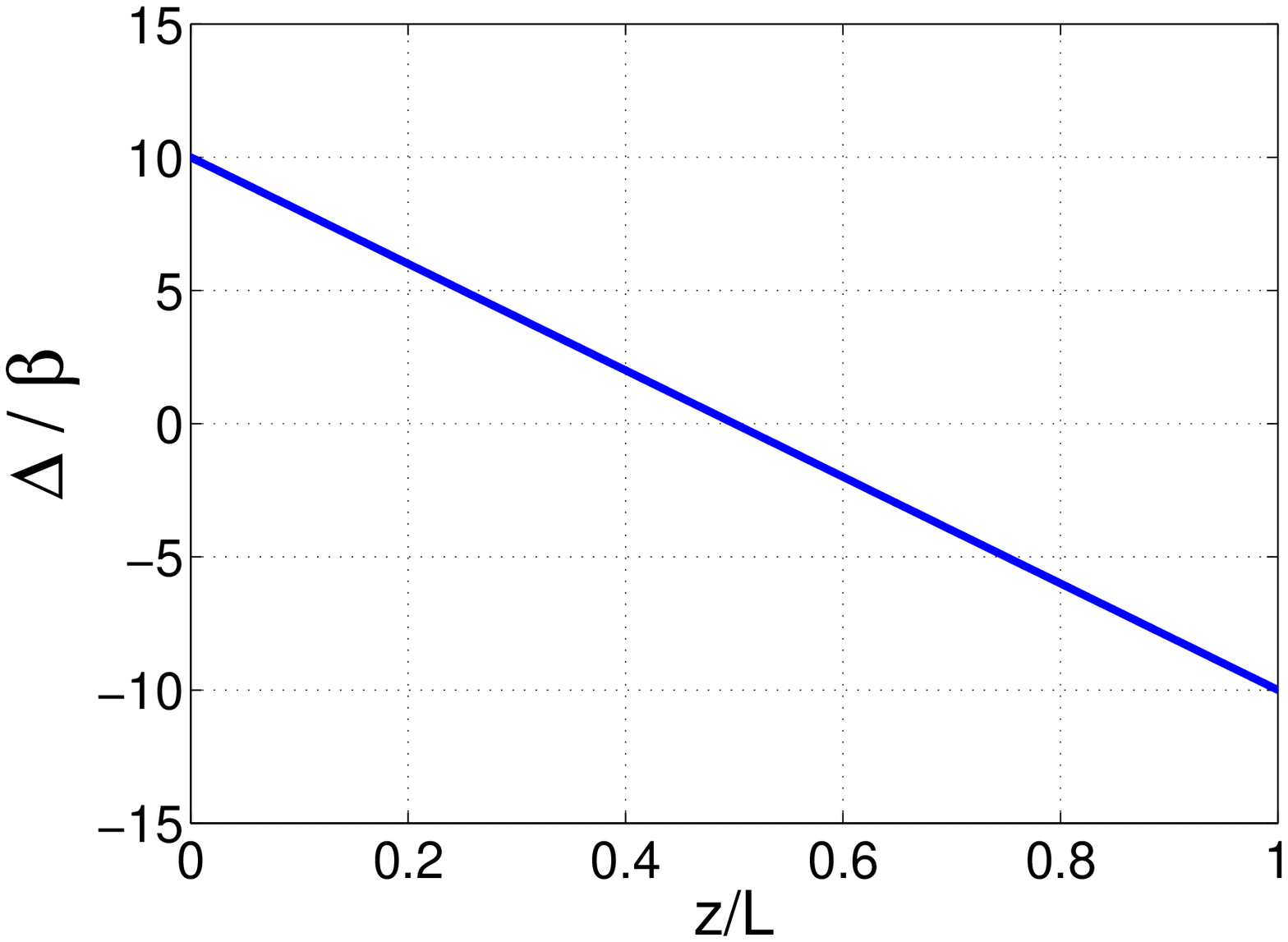}}

\label{Comparing_GEM_Delta}
\begin{singlespace}
\small{
\caption{(a) Comparison of the AFS memory output (green) with GEM output (red) when we send the same input pulse (blue). (b) illustrates the detuning as a function of time in AFS memory. (c), (d) show the detuning as a function of spatial coordinate respectively for storage and retrieval in GEM memory. In the simulation of GEM, at time t=1.6L/c the detuning of the atoms is flipped. The effective spatial gradient ($\frac{\dot{\Delta}}{c}$) in the AFS memory is set equal to the spatial gradient in GEM for both storage and retrieval.}}
\end{singlespace}
\end{figure}

We have also performed numerical comparisons between GEM \cite{GEM_Analytical} and AFS, where in the case of AFS the effective spatial gradient for light propagating at velocity $c$ ($\frac{\dot{\Delta}}{c}$) is set equal to the spatial gradient for GEM. Figure 8(a) shows the agreement between outputs of the two protocols. This connection can be captured rigorously by going to the retarded frame ($\tau\rightarrow t-z/c$, $z'=z$). In the retarded frame the time-dependent detuning $\Delta(t)$ is transformed to $\Delta(\tau + z/c)$, which depends on both retarded time and space. However, for pulses smaller than the medium, the term $\tau$ can approximately be
neglected compared to $z/c$. As a result we end up with a space-dependent detuning $\Delta(z/c)$.

It is important to note that this connection is valid under the assumption that retarded time is much smaller than the temporal extension of medium ($L/c$), which is true for the short pulses traveling with almost the speed of light. However based on the polaritonic description we know that the pulse slows down in the medium, and as a result the retarded time increases more than the time duration of the pulse and our assumption is no longer valid. What justifies our approximation is that, when the retarded time exceeds $L/c$, the pulse is almost absorbed. Thus this factor doesn't play an important role. The small discrepancy of the two protocols in figure 8 can be explained by the imperfection of this approximation.

\end{document}